\documentclass[fleqn,usenatbib]{mnras}

\usepackage{newtxtext,newtxmath}
\usepackage[T1]{fontenc}

\DeclareRobustCommand{\VAN}[3]{#2}
\let\VANthebibliography\thebibliography
\def\thebibliography{\DeclareRobustCommand{\VAN}[3]{##3}\VANthebibliography}

\usepackage{graphicx}	% Including figure files
\usepackage{amsmath}	% Advanced maths commands
\defcitealias{Tang2019}{T19}

\title[Stellar populations of EELGs]{Stellar populations and star formation histories of the most extreme [OIII] emitters at $\mathbf{z=1.3-3.7}$}

\author[M. Tang et al.]{
Mengtao Tang$^{1}$\thanks{mengtao.tang@ucl.ac.uk}, 
Daniel P. Stark$^{2}$ and 
Richard S. Ellis$^{1}$
\\
\\
$^{1}$ Department of Physics and Astronomy, University College London, Gower Street, London WC1E 6BT, UK \\
$^{2}$ Steward Observatory, University of Arizona, 933 N Cherry Ave, Tucson, AZ 85721, USA \\
}

\date{Accepted XXX. Received YYY; in original form ZZZ}

\pubyear{2022}

\begin{document}
\label{firstpage}
\pagerange{\pageref{firstpage}--\pageref{lastpage}}
\maketitle

\begin{abstract}
As the {\it James Webb Space Telescope} approaches scientific operation, there is much interest in exploring the redshift range beyond that accessible with {\it Hubble Space Telescope} imaging. Currently, the only means to gauge the presence of such early galaxies is to age-date the stellar population of systems in the reionisation era. As a significant fraction of $z\simeq7-8$ galaxies are inferred from {\it Spitzer} photometry to have extremely intense [O~{\small III}] emission lines, it is commonly believed these are genuinely young systems that formed at redshifts $z<10$, consistent with a claimed rapid rise in the star formation density at that time. Here we study a spectroscopically-confirmed sample of extreme [O~{\small III}] emitters at $z=1.3-3.7$, using both dynamical masses estimated from [O~{\small III}] line widths and rest-frame UV to near-infrared photometry to illustrate the dangers of assuming such systems are genuinely young. For the most extreme of our intermediate redshift line emitters, we find dynamical masses $10-100$ times that associated with a young stellar population mass, which are difficult to explain solely by the presence of additional dark matter or gaseous reservoirs. Adopting nonparametric star formation histories, we show how the near-infrared photometry of a subset of our sample reveals an underlying old ($>100$~Myr) population whose stellar mass is $\simeq40$ times that associated with the starburst responsible for the extreme line emission. Without adequate rest-frame near-infrared photometry we argue it may be premature to conclude that extreme line emitters in the reionisation era are low mass systems that formed at redshifts below $z\simeq10$.
\end{abstract}

\begin{keywords}
cosmology: observations - galaxies: evolution - galaxies: formation - galaxies: high-redshift
\end{keywords}

%%%%%%%%%%%%%%%%%%%%%%%%%%%%%%%%%%%%%%%%%%%%%%%%%%

%%%%%%%%%%%%%%%%% BODY OF PAPER %%%%%%%%%%%%%%%%%%

%%%%%%%%%%%% INTRODUCTION %%%%%%%%%%%%

\section{Introduction} \label{sec:intro}

Following the successful launch of the {\it James Webb Space Telescope (JWST)}, there is increased interest in exploring the cosmic era beyond the redshift $z\simeq10-11$ horizon established via deep imaging of blank and gravitationally-lensed fields with the {\it Hubble Space Telescope (HST)} \citep[e.g.,][]{Ellis2013,Oesch2016,Salmon2018,Jiang2021}. The census of star-forming galaxies revealed during the reionisation era delineates a continuous decline with increasing redshift over $7<z<10$ \citep[e.g.,][]{McLeod2016} with possible evidence of a more rapid assembly prior to a redshift $z\simeq8$ \citep[e.g.,][]{Oesch2014,Oesch2018}. Such trends have been claimed to indicate the onset of reionisation at $z\simeq10-12$ is consistent with electron scattering measures of the microwave background \citep[e.g.,][]{Robertson2015,Planck2020}.

Independent verification of the early cosmic star formation history might be obtained from the stellar ages of the most distant galaxies. Limited spectrophotometric data for a few $z\simeq9$ galaxies, where {\it Spitzer}/Infrared Array Camera (IRAC) photometry is free from nebular emission line contamination indicates the possibility of star formation beyond $z\simeq12$ \citep{Hashimoto2018,Roberts-Borsani2020,Laporte2021}. But this inference relies on the assumed past star formation history and thus remains uncertain. Such early activity is also hard to reconcile with the observation that many galaxies in the redshift interval $6.6<z<9$ have prominent ``IRAC excesses'' most easily explained by intense [O~{\small III}]+H$\beta$ line emission indicative of young ($\simeq10$~Myr) stellar populations \citep[e.g.,][]{Labbe2013,Smit2014,Smit2015,Roberts-Borsani2016,Endsley2021,Stefanon2022}. In the latter case, however, the question remains as to whether this strong line emission is the result of an energetic phase of secondary star formation which could mask the presence of an older stellar population. Unfortunately the depth of the Spitzer/IRAC photometry in its longest wavelength passbands at $5.7$ and $7.9\ \mu$m (equivalent to rest-frame $6300-8900$~\AA\ at $z=8$) is insufficient to address this possibility for individual galaxies in the reionisation era.

The star formation history of such ``extreme emission line galaxies'' (EELGs) is best addressed through detailed studies of lower redshift analogues where suitably deep rest-frame optical and near-infrared (NIR) photometry of individual examples is available. Sizable samples of $z\simeq1-2$ galaxies with large [O~{\small III}]+H$\beta$ equivalent widths (EWs) have been identified in broadband imaging and spectroscopic surveys \citep[e.g.][]{Atek2011,Atek2014,vanderWel2011,Maseda2014,Amorin2015} revealing that these galaxies are low mass systems ($M_{\star}\simeq10^8-10^9\ M_{\odot}$) undergoing bursts of star formation (age $\simeq10-100$~Myr). In \citet[][hereafter \citetalias{Tang2019}]{Tang2019}, we built on these studies, investigating the rest-frame optical spectra of over $200$ extreme [O~{\small III}] emitting galaxies at $z\simeq1-2$ with [O~{\small III}]~$\lambda5007$ EW $>225$~\AA. In particular, we targeted $\simeq30$ of the most extreme optical line emitters with EW$_{\rm{[OIII]}\lambda5007}>800$~\AA. Although such a population is rare at $z\sim2$ \citep{Boyett2021}, it is quite typical in the reionisation era \citep{Endsley2021}. In \citetalias{Tang2019}, we demonstrated that the most extreme [O~{\small III}] emitters are dominated by very young stellar populations with age $<10$~Myr (assuming a constant star formation history). However, we did not negate the possibility of an older stellar populations whose presence might be masked by a younger starburst. 

In this paper we aim to constrain the presence of evolved stellar populations in the most extreme [O~{\small III}] emitting galaxies. We will address this question using two complementary probes: dynamical masses derived from gaseous line widths, and star formation histories (SFHs) derived by fitting rest-frame UV to near-infrared (NIR) photometry. If older populations ($>100$~Myr up to a few Gyr) contribute significantly (in stellar mass) to these systems, we would expect to see a very large dynamical mass compared to the stellar mass of the young stellar population and, furthermore, we would expect radiation from the older stars to be detectable in the rest-frame NIR photometry. Although obtaining such detailed information is not currently practical for EELGs in the reionisation era, our goal is to use our low redshift EELG analogues to illustrate the possibility that the ages of such $z\simeq7-8$ galaxies may have been significantly underestimated, and thus their presence may be consistent with star formation to redshifts beyond $z\simeq10-12$.

A plan of the paper follows. In Section~\ref{sec:OA} we introduce the sample of $z=1.3-3.7$ extreme [O~{\small III}] emitters drawn from \citetalias{Tang2019} and define two subsamples for which we have secured, for the first case, velocity dispersions and dynamical masses from high-resolution spectra and, for the second case, spectral energy distributions (SEDs) extending from the rest-frame UV to the NIR. For the latter subsample we derive physical properties such as stellar masses, ages and star-formation rates from the SEDs in Section~\ref{sec:fitting}. By contrasting the stellar and dynamical masses in the context of the EW$_{\rm{[OIII]}\lambda5007}$, we present new evidence for evolved stellar populations in the most extreme line emitters in Section~\ref{sec:old_star}. Finally, we discuss the implications of our findings for similar sources in the reionisation era in Section~\ref{sec:discussion}. We adopt a $\Lambda$-dominated, flat universe with $\Omega_{\Lambda}=0.7$, $\Omega_{\rm{M}}=0.3$, and $H_0=70$~km~s$^{-1}$~Mpc$^{-1}$. All magnitudes in this paper are quoted in the AB system \citet{Oke1983}, and all EWs are quoted in the rest frame.

%%%%%%%%%%%% OBSERVATION AND ANALYSIS %%%%%%%%%%%%

\section{Observations and Analysis} \label{sec:OA}

To derive the dynamical masses of EELGs, we measure velocity dispersions from the high-resolution ($R>3000$) spectra obtained via the Multi-object Spectrometer for Infrared Exploration (MOSFIRE; \citealt{McLean2010,McLean2012}) on the Keck telescope, which is a part of our large NIR (rest-frame optical) spectroscopic survey of extreme [O~{\small III}] emitters at $z=1.3-3.7$ \citepalias{Tang2019}. We also select a subset of the most extreme [O~{\small III}] emitters with robust mid-infrared (rest-frame NIR) photometry measurements from our spectroscopic sample in \citetalias{Tang2019}. In this section, we briefly summarize our spectroscopic survey (Section~\ref{sec:sample}), and describe the data analysis and the samples used in this paper (Section~\ref{sec:analysis}).

\subsection{Rest-frame optical spectroscopy of extreme [O~{\small III}] emitters at $\mathbf{z=1.3-3.7}$} \label{sec:sample}

The dataset studied in this work is taken from our large rest-frame optical spectroscopic survey of extreme [O~{\small III}] emitting galaxies at $z=1.3-3.7$ in the Cosmic Assembly Near-infrared Deep Extragalactic Legacy Survey (CANDELS; \citealt{Grogin2011,Koekemoer2011}) fields. We direct the reader to \citetalias{Tang2019} for the full description of the sample selection and the follow-up spectroscopic observations of this survey. In brief, the EELGs were identified based on the [O~{\small III}] EWs inferred from 3D-HST \citep{Brammer2012,Skelton2014,Momcheva2016} grism spectra (at $z=1.3-2.4$; \citetalias{Tang2019}) or the {\it K}-band flux excess (at $z=3.1-3.7$; Tang et al., in preparation). We require the emitters to have rest-frame [O~{\small III}]~$\lambda\lambda4959,5007$ EWs $=300-3000$~\AA, which match values inferred to be common in reionisation-era systems \citep[e.g.,][]{Endsley2021}. We obtain NIR spectra with the MMT and Magellan Infrared Spectrograph (MMIRS; \citealt{McLeod2012,Chilingarian2015}) on the MMT and Keck/MOSFIRE, targeting strong rest-frame optical emission lines ([O~{\small II}], [Ne~{\small III}], H$\beta$, [O~{\small III}], and H$\alpha$). 

In \citetalias{Tang2019}, we presented NIR spectra of $227$ EELGs obtained between the 2015B and 2018A semesters. Between the 2018B and 2019B semesters, we continued our NIR spectroscopic survey, acquiring rest-frame optical spectra for an additional $68$ EELGs at $z=1.3-3.7$ following the same observing strategy described in \citetalias{Tang2019}. Spectra of $31$ of these $68$ targets were obtained using MMT/MMIRS in the 2018B and 2019B semesters. We have collected $24$ hours of on-source integration, targeting on the Ultra Deep Survey (UDS) field with three separate multi-object slit masks. MMIRS spectra were taken with the $J$ grism + $zJ$ filter, $H3000$ grism + $H$ filter, and $K3000$ grism + $Kspec$ filter sets with a slit width of $1$ arcsec for science targets. The $1$ arcsec slit width with MMIRS results in a resolving power of $R\approx1000$. The average seeing during observations was between $0.8$ and $1.5$ arcsec.

Spectra of the remaining $37$ targets were obtained using Keck/MOSFIRE on 2019 April 15 and 16. We targeted on the All-Wavelength Extended Groth Strip International Survey (AEGIS) and the Great Observatories Origins Deep Survey North (GOODS-N) fields with three multi-object slit masks with a total on-source integration time of $13.6$ hours. The MOSFIRE masks were primarily focused on $z\gtrsim9$ galaxies \citep{Laporte2021}, and EELGs at lower redshift were placed as fillers. Spectra were taken in the $J$ band with a slit width of $0.7$ arcsec, which results in a resolution of $R=3318$. This resolution allows us to resolve the strong [O~{\small III}]~$\lambda5007$ emission lines in the wavelength direction and measure the velocity dispersion (Section~\ref{sec:analysis}). The average seeing during the MOSFIRE observation was between $0.7$ and $1.1$ arcsec. 

We reduced the MMIRS and MOSFIRE spectra using the public available data reduction pipelines for the two instruments\footnote{MMIRS: \url{https://bitbucket.org/chil_sai/mmirs-pipeline};\\ MOSFIRE: \url{https://keck-datareductionpipelines.github.io/MosfireDRP}}. These pipelines perform flat-fielding, wavelength calibration, and background subtraction before 2D spectra extraction. The 1D spectra extraction and flux calibration were performed following the methods described in \citetalias{Tang2019}. We created 1D spectra from the reduced 2D spectra using a boxcar extraction. The telluric absorption and instrumental response were determined using observations of A0V stars. Slit loss correction of each target was performed using the in-slit light fraction computed from its {\it HST} image following the procedures described in \citet{Kriek2015}. We then performed the absolute flux calibration using observations of slit stars, by comparing the  slit-loss corrected count rates of slit star spectra with the broadband flux in the \citet{Skelton2014} catalogues. Details of the observations between 2018B and 2019B are summarized in Table~\ref{tab:obs}.

%%%% Table: Summary of new observations %%%%

\begin{table*}
\begin{tabular}{|c|c|c|c|c|c|c|c|c|c|}
\hline
Instrument & Mask Name & Number of Target & R.A. & Decl. & P.A. & Grism & Filter & Exposure Time & Average Seeing \\
 & & & (hh:mm:ss) & (dd:mm:ss) & (deg) & & & (seconds) & ($''$) \\
(1) & (2) & (3) & (4) & (5) & (6) & (7) & (8) & (9) & (10) \\
\hline
MMT/MMIRS & udse04 & 22 & 2:17:37.000 & $-$5:11:27.00 & $-$97.00 & \textit{J} & \textit{zJ} & 14400 & 0.8 \\
MMT/MMIRS & udse04 & 22 & 2:17:37.000 & $-$5:11:27.00 & $-$97.00 & \textit{H3000} & \textit{H} & 14400 & 1.1 \\
MMT/MMIRS & udse04 & 22 & 2:17:37.000 & $-$5:11:27.00 & $-$97.00 & \textit{K3000} & \textit{Kspec} & 10800 & 1.5 \\
MMT/MMIRS & udse05 & 15 & 2:17:15.000 & $-$5:13:45.00 & 95.00 & \textit{J} & \textit{zJ} & 14400 & 1.0 \\
MMT/MMIRS & udse05 & 15 & 2:17:15.000 & $-$5:13:45.00 & 95.00 & \textit{H3000} & \textit{H} & 7200 & 0.8 \\
MMT/MMIRS & udse07 & 16 & 2:17:11.100 & $-$5:13:47.00 & $-$99.00 & \textit{H3000} & \textit{H} & 14400 & 0.8 \\
MMT/MMIRS & udse07 & 16 & 2:17:11.100 & $-$5:13:47.00 & $-$99.00 & \textit{K3000} & \textit{Kspec} & 10800 & 0.8 \\
Keck/MOSFIRE & EGSY2\_1 & 13 & 14:19:56.56 & $+$52:54:22.02 & 130.0 & \textit{J} & - & 9600 & 0.8 \\
Keck/MOSFIRE & GNz9\_1b & 9 & 12:37:06.71 & $+$62:17:42.90 & 142.0 & \textit{J} & - & 20160 & 0.7 \\
Keck/MOSFIRE & GNz10\_1 & 15 & 12:36:25.45 & $+$62:14:39.60 & 230.0 & \textit{J} & - & 19200 & 1.1 \\
\hline
\end{tabular}
\caption{Summary of the NIR spectroscopic observations between 2018B and 2019B semesters. Totally $37$ targets were placed on three Keck/MOSFIRE masks. And $31$ individual targets were placed on three MMT/MMIRS masks, including $22$ targets being placed on more than one mask in order to get multiple strong rest-frame optical emission lines. Column (1): telescope and instrument used; Column (2): mask name; Column (3): number of science targets on each mask, alignment stars and slit stars are not included; Column (4): right ascension of the mask center; Column (5): declination of the mask center; Column (6): position angle of the mask; Column (7): grism of the mask observed; Column (8): filter of the mask observed; Column (9): Total exposure time of the mask in each grism + filter set; Column (10): average seeing during the observation.}
\label{tab:obs}
\end{table*}

\subsection{Data analysis and sample selection} \label{sec:analysis}

The emission line measurements of the spectra taken from 2018B to 2019B were performed using the same procedures described in \citetalias{Tang2019}. We have confirmed redshifts of $64$ extreme [O~{\small III}] emitters in this data set. In the remaining $4$ objects for which we fail to measure redshifts, either the spectra have very low S/N or the emission lines are contaminated by sky line residuals. The emission line fluxes were measured by fitting Gaussian profiles to the lines in the 1D spectra. The nebular gas extinction $E(B-V)$ is computed by comparing the observed H$\alpha$/H$\beta$ ratio (Balmer decrement) to the intrinsic value $2.86$ \citep{Osterbrock2006} and assuming the \citet{Cardelli1989} extinction curve. Using the line fluxes and the underlying continuum inferred from the best-fitting SEDs\footnote{ Because the S/N of the underlying continuum measured from spectra is usually low, we adopt the continuum inferred from the best-fitting SEDs which provides an improved determination of the continuum.} (Section~\ref{sec:fitting}), we calculate the EWs of [O~{\small II}], H$\beta$, [O~{\small III}], and H$\alpha$ emission lines. Together with the $227$ NIR spectra  previously taken, we have now constructed a rest-frame optical spectroscopic sample of $291$ extreme [O~{\small III}] emitters at $z=1.3-3.7$.

One of the goals of this study is to estimate the dynamical masses of EELGs at $z\sim2$. We follow the procedures in \citet{Maseda2013} to derive the dynamical mass, using the velocity dispersion measured from the width of the [O~{\small III}]~$\lambda5007$ emission line (i.e., the most luminous rest-frame optical emission line with the highest S/N in our sample) and the effective radius measured from {\it HST} imaging. To measure the velocity dispersion, the spectral resolution must be sufficient to deconvolve the intrinsic line width from the observed width, which can only be done with Keck/MOSFIRE spectra ($R=3300-3700$) in our spectroscopic sample. Therefore, we select a subsample of EELGs with Keck/MOSFIRE observations from our sample, which were taken in three observing runs (2015 November and 2016 April, \citetalias{Tang2019}, and 2019 April). In total there are $59$ sources with MOSFIRE spectra revealing [O~{\small III}]~$\lambda5007$ emission lines. To robustly measure the line width, we exclude objects with low S/N ($<5$) line measurements or emission lines contaminated by sky line residuals. We also remove sources that are likely interacting systems, including galaxies showing nearby counterparts or irregular morphologies, which would otherwise influence on both the emission line width and the radius measurements \citep[e.g.,][]{Price2016}. By visually inspecting the images, $25$ out of the $59$ galaxies were removed from the sample. As a result, the subsample used to estimate dynamical masses contains $34$ EELGs at $z=1.3-2.4$ (hereafter Sample I). The [O~{\small III}]~$\lambda5007$ EWs of objects in Sample I are $100-1000$~\AA, covering the EW range of typical $z\sim7$ galaxies \citep[e.g.,][]{Endsley2021}.

\begin{figure*}
\begin{center}
\includegraphics[width=0.49\linewidth]{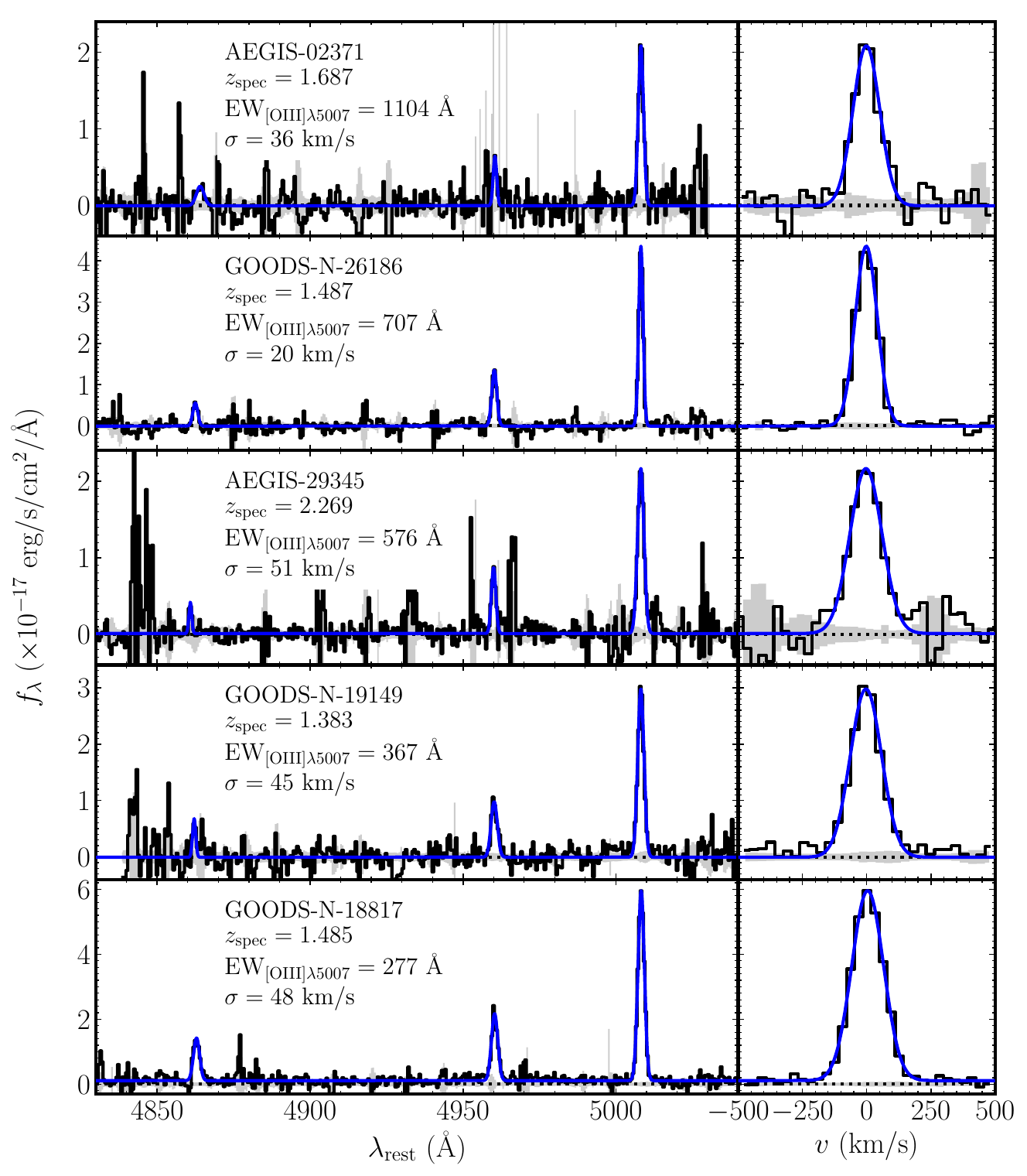}
\includegraphics[width=0.49\linewidth]{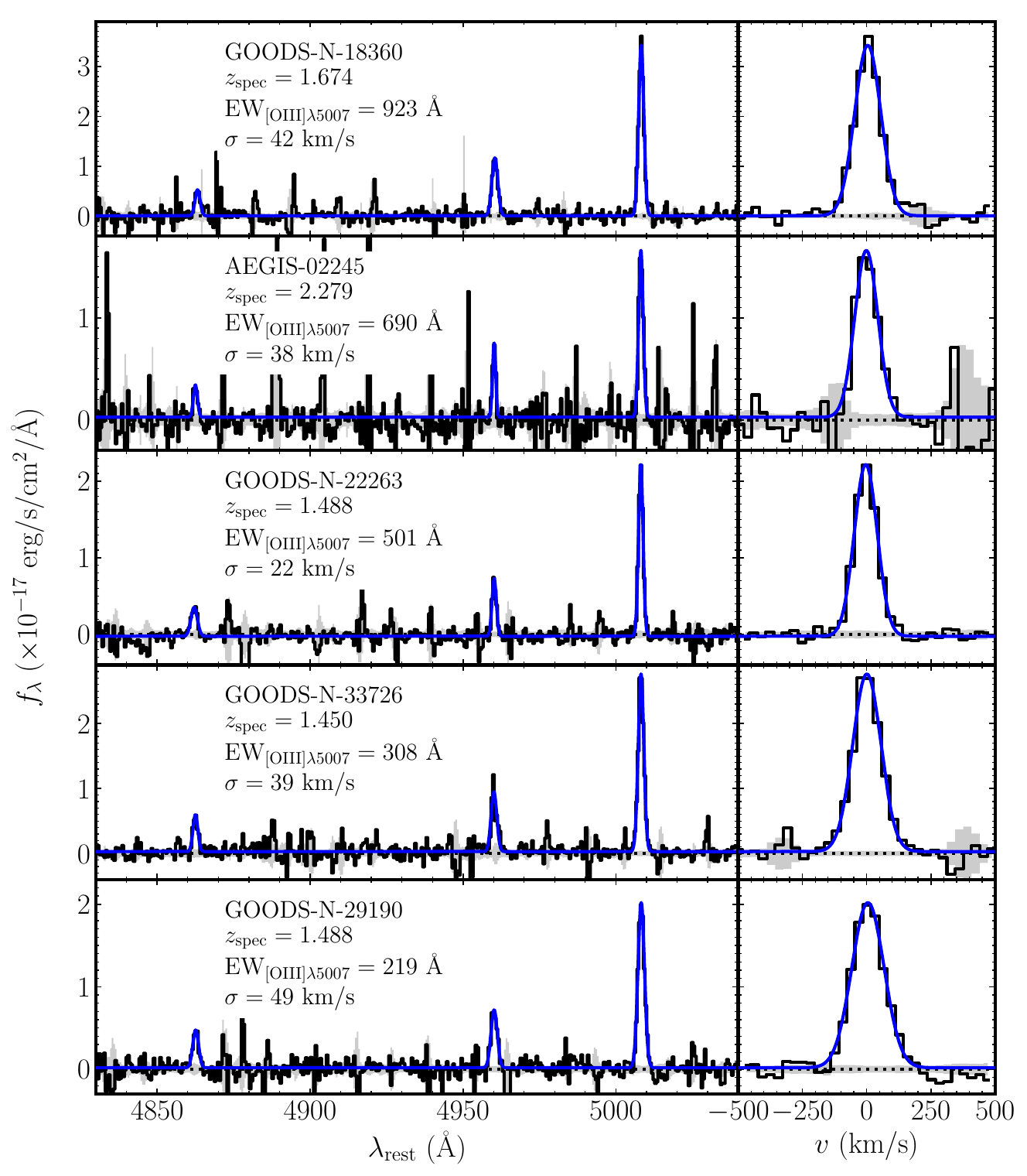}
\caption{Example Keck/MOSFIRE spectra of ten EELGs at $z=1.3-2.4$ in our Sample I, in decreasing [O~{\scriptsize III}] EW order. The left panel of each plot shows detections of H$\beta$, [O~{\scriptsize III}]~$\lambda4959$, and [O~{\scriptsize III}]~$\lambda5007$ emission lines. Blue curves present the best-fitting emission line profiles, and grey shaded regions present $\pm1\sigma$ uncertainties. The right panel of each plot shows the zoom-in [O~{\scriptsize III}]~$\lambda5007$ emission line profile (in velocity space), and the line width is used to compute the velocity dispersion. The resolution of our Keck/MOSFIRE spectra ranges from $R=3318$ to $R=3660$.}
\label{fig:spec}
\end{center}
\end{figure*}

\begin{figure}
\begin{center}
\includegraphics[width=\linewidth]{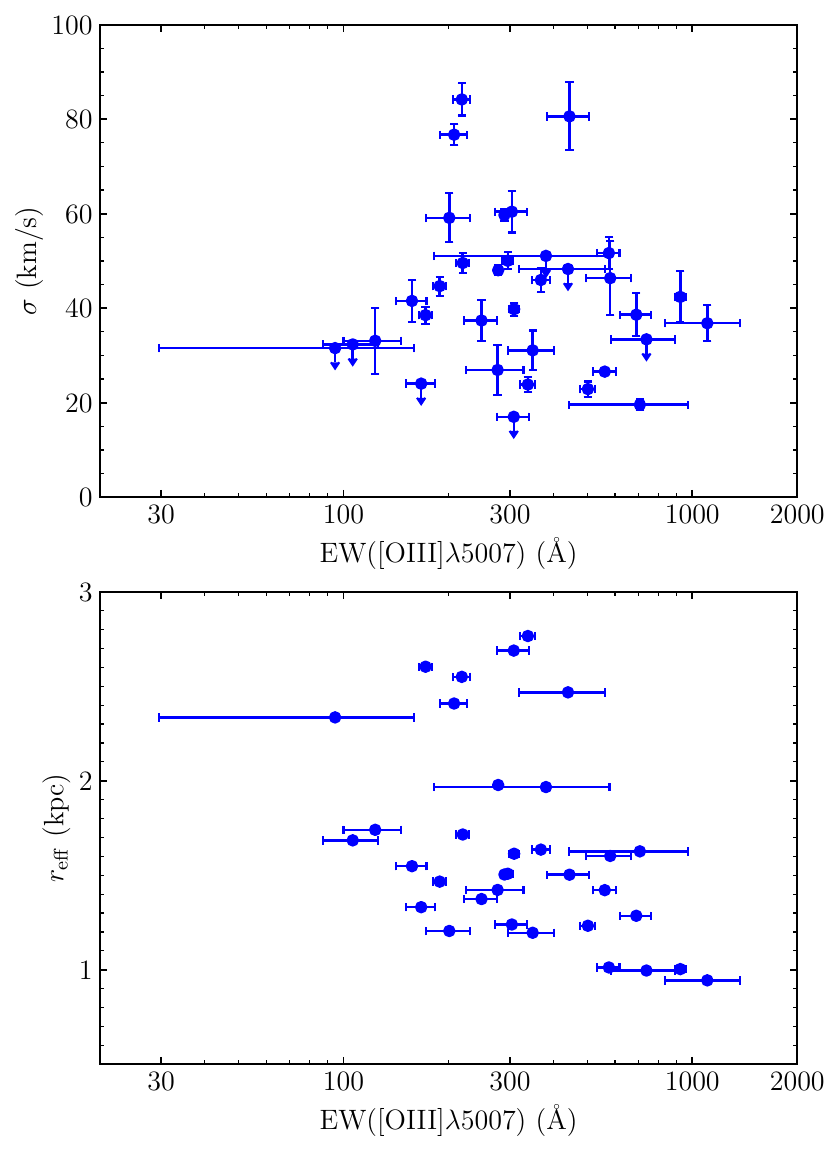}
\caption{The velocity dispersion (measured from [O~{\scriptsize III}]~$\lambda5007$ emission line width; top panel) and the effective radius (half-light radius; bottom panel) as functions of [O~{\scriptsize III}]~$\lambda5007$ EW for our EELGs at $z=1.3-2.4$ with Keck/MOSFIRE spectra (Sample I). Objects with velocity dispersion smaller than the instrument resolution are shown as $3\sigma$ upper limits. Galaxies with the largest [O~{\scriptsize III}] EWs tend to have smaller velocity dispersions, though with large scatter. Galaxies with larger [O~{\scriptsize III}] EWs are also more compact sources with smaller sizes.}
\label{fig:vdis_reff}
\end{center}
\end{figure}

To derive the velocity dispersions of objects in Sample I, we compute the intrinsic [O~{\small III}]~$\lambda5007$ line width by subtracting the instrument resolution in quadrature from the observed line widths. The observed line width is derived from fitting the [O~{\small III}]~$\lambda5007$ emission line with a Gaussian function. In Fig.~\ref{fig:spec} we show examples of the Keck/MOSFIRE spectra and [O~{\small III}]~$\lambda5007$ profiles of objects in Sample I. The resulting velocity dispersions ($\sigma$) of the $34$ sources are in the range $20-84$~km~s$^{-1}$, with a median value of $42$~km~s$^{-1}$. We find that all the [O~{\small III}]~$\lambda5007$ lines can be well fit by single Gaussian profiles with no evident of additional broader components (e.g., $\sigma>130$~km~s$^{-1}$) driven by outflows \citep[e.g.,][]{Newman2012,ForsterSchreiber2014,Freeman2019}. In the top panel of Fig.~\ref{fig:vdis_reff}, we plot the velocity dispersion of Sample I as function of the [O~{\small III}] EW finding a very weak correlation with the nonparametric Spearman rank correlation coefficient $\rho=-0.25$ and $p$-value $p=0.21$. This is in the sense that the most extreme [O~{\small III}] emitters tend to have smaller velocity dispersions. The velocity dispersions of our EELGs are smaller than those of more massive ($M_{\star}\sim10^{10}\ M_{\odot}$) star-forming galaxies at $z\sim2$ selected from rest-frame UV colors ($\langle\sigma\rangle=108$~km~s$^{-1}$; \citealt{Erb2006}) or rest-frame optical magnitude (median $\sigma=78$~km~s$^{-1}$; \citealt{Price2016}). The velocity dispersions of our Sample I are also slightly smaller than the values of $z\sim1-2$ EELGs in \citet[][$\sigma=53$~km~s$^{-1}$]{Maseda2014}, which are $\sim0.6$~mag brighter (median $m_{\rm{F606W}}=24.9$) than our sources (median $m_{\rm{F606W}}=25.5$).

We also measure the effective radii of the objects in Sample I. Here we use the half-light radii (in pixels) provided by \citet{Skelton2014} catalogues, which are measured from {\it HST}/WFC3 F160W images by using SExtractor \citep{Bertin1996} and adopt these as virial radii \citep[e.g.,][]{Maseda2013}. The effective radii of the EELGs in Sample I range from $0.9$ kpc to $2.8$ kpc, with a median value of $1.5$ kpc. As these are larger than the half width at half maximum of the point spread function of F160W imaging, the sources are adequately resolved. In the bottom panel of Fig.~\ref{fig:vdis_reff}, we show the effective radius as functions of [O~{\small III}] EW. The two quantities show a moderate correlation with the Spearman correlation coefficient $\rho=-0.43$ and $p$-value $p=1.0\times10^{-2}$, and it is clear that galaxies with more extreme optical line emission are more compact. The physical properties of the EELGs in Sample I are summarised in Table~\ref{tab:dyn_sample}.

In order to constrain the stellar populations and SFHs of the most extreme [O~{\small III}] emitting galaxies (EW$_{\rm{[OIII]}\lambda5007}>800$~\AA), we select a second subsample of objects with robust rest-frame UV-to-NIR photometry measurements from our spectroscopic sample. In \citetalias{Tang2019}, we demonstrated that galaxies with the largest optical line EWs likely undergo recent bursts of star formation ($<10$~Myr, assuming constant CSFH). The strong nebular continuum and line emission reprocessed by the radiation fields emitted from very young stars dominates the rest-frame UV-to-optical SEDs and may obscure the light from much older stellar populations. However, stars older than a few hundred Myr would be more dominant at the rest-frame NIR wavelengths, and we aim to constrain the potential older stellar populations with the rest-frame UV-to-NIR SEDs. At $z=1.3-3.7$, the rest-frame NIR fluxes have been shifted to mid-infrared (MIR), which can be probed by {\it Spitzer}/IRAC $3.6\ \mu$m and $4.5\ \mu$m photometry. Therefore, we select a subsample of galaxies with [O~{\small III}]~$\lambda5007$ EW $>800$~\AA\ and high S/N ($>5$) [O~{\small III}] and H$\alpha$ emission line measurements (to better constrain the nebular emission at rest-frame optical wavelengths), containing robust IRAC detections. 

Due to the relatively low resolution of the {\it Spitzer} images, contamination from neighbouring objects to the target needs to be taken into account when determining the robust IRAC flux. \citet{Skelton2014} used the high-resolution {\it HST} image as a prior to estimate and subtract the contribution from neighbouring blended sources in the low-resolution {\it Spitzer} image. In order to minimize the effect of neighbouring contamination, we adopt a S/N $>5$ selection for IRAC $3.6\ \mu$m and $4.5\ \mu$m measurements, and restrict the ratio of contaminating flux to be $<0.5$. In this manner we select a subsample of $7$ extreme [O~{\small III}] emitting galaxies at $z=1.3-3.7$ (hereafter Sample II). Their physical properties are presented in Table~\ref{tab:irac_sample}.

Finally, we exclude the possibility that the IRAC fluxes of the objects in Sample II arise from active galactic nucleus (AGN) activity. In our spectroscopic sample of EELGs, we already removed sources that are likely host X-ray AGN \citepalias{Tang2019}. We can test whether the IRAC fluxes are consistent with the presence of an AGN using the selection criteria adopted by \citet{Donley2012} to identify IR AGN at $z\sim1-3$ \citep[e.g.,][]{Coil2015}. These criteria exploit the fact that IR AGN tend to have red IRAC SEDs (see Equation 1 and 2 in \citealt{Donley2012}) and we find that none of the emitters in our Sample II have IRAC colors consistent with the \citet{Donley2012} criteria. 

%%%% Figure: Keck/MOSFIRE spectra %%%%

%%%% Figure: Velocity dispersion vs. [OIII] EW & Effective radius vs. [OIII] EW %%%%

%%%% Table: Properties of EELGs in Sample I %%%%

\begin{table*}
\begin{tabular}{|c|c|c|c|c|c|c|c|c|c|}
\hline
ID & R.A. & Decl. & $z_{\rm{spec}}$ & EW$_{\rm{[OIII]}\lambda5007}$ & $\log{(M_{\star,\rm{CSFH}}/M_{\odot})}$ & sSFR$_{\rm{CSFH}}$ & $\sigma$ & $r_{\rm{eff}}$ & $\log{(M_{\rm{dyn}}/M_{\odot})}$ \\
 & (hh:mm:ss) & (dd:mm:ss) & & (\AA) & & (Gyr$^{-1}$) & (km~s$^{-1}$) & (kpc) & \\
\hline
COSMOS-19180 & 10:00:26.847 & +02:22:26.727 & $1.213$ & $295\pm10$ & $8.74^{+0.05}_{-0.06}$ & $13^{+2}_{-2}$ & $50\pm1$ & $1.5$ & $9.42\pm0.38$ \\
GOODS-S-28288 & 03:32:18.251 & -27:46:51.964 & $1.234$ & $188\pm 8$ & $9.02^{+0.05}_{-0.05}$ & $3^{+0}_{-0}$ & $44\pm2$ & $1.5$ & $9.31\pm0.38$ \\
UDS-27523 & 02:17:06.812 & -05:11:00.694 & $1.670$ & $123\pm23$ & $9.36^{+0.08}_{-0.06}$ & $2^{+0}_{-0}$ & $33\pm7$ & $1.7$ & $9.12\pm0.44$ \\
UDS-36954 & 02:17:14.900 & -05:09:06.174 & $1.658$ & $248\pm27$ & $9.12^{+0.08}_{-0.07}$ & $6^{+2}_{-1}$ & $37\pm4$ & $1.4$ & $9.13\pm0.40$ \\
UDS-37070 & 02:17:04.624 & -05:09:05.512 & $1.416$ & $ 94\pm64$ & $9.61^{+0.05}_{-0.06}$ & $1^{+0}_{-0}$ & $<31$ & $2.3$ & $<9.21$ \\
AEGIS-02245 & 14:20:14.359 & +52:54:09.481 & $2.279$ & $690\pm70$ & $8.19^{+0.20}_{-0.19}$ & $38^{+21}_{-16}$ & $38\pm4$ & $1.3$ & $9.12\pm0.40$ \\
AEGIS-14784 & 14:20:08.796 & +52:56:21.812 & $2.291$ & $218\pm12$ & $9.76^{+0.07}_{-0.06}$ & $11^{+3}_{-2}$ & $84\pm3$ & $2.5$ & $10.10\pm0.38$ \\
AEGIS-15929 & 14:20:05.999 & +52:56:10.029 & $2.206$ & $444\pm62$ & $8.67^{+0.12}_{-0.14}$ & $33^{+24}_{-12}$ & $80\pm7$ & $1.5$ & $9.83\pm0.39$ \\
AEGIS-17167 & 14:19:55.518 & +52:54:36.796 & $2.207$ & $157\pm15$ & $9.61^{+0.09}_{-0.15}$ & $4^{+2}_{-1}$ & $41\pm4$ & $1.5$ & $9.27\pm0.39$ \\
AEGIS-29345 & 14:19:49.797 & +52:56:30.463 & $2.269$ & $576\pm41$ & $8.15^{+0.22}_{-0.10}$ & $70^{+16}_{-32}$ & $51\pm3$ & $1.0$ & $9.27\pm0.38$ \\
AEGIS-02371 & 14:20:47.930 & +53:00:06.537 & $1.687$ & $1104\pm268$ & $7.18^{+0.17}_{-0.12}$ & $151^{+57}_{-49}$ & $36\pm3$ & $0.9$ & $8.95\pm0.39$ \\
AEGIS-17916 & 14:20:25.737 & +53:00:08.473 & $1.628$ & $307\pm31$ & $9.06^{+0.06}_{-0.08}$ & $4^{+1}_{-0}$ & $<17$ & $2.7$ & $<8.73$ \\
AEGIS-10988 & 14:20:02.853 & +52:54:26.496 & $1.566$ & $106\pm19$ & $8.91^{+0.19}_{-0.19}$ & $1^{+1}_{-0}$ & $<32$ & $1.7$ & $<9.09$ \\
AEGIS-15240 & 14:19:56.598 & +52:54:16.966 & $1.648$ & $171\pm 7$ & $9.25^{+0.08}_{-0.09}$ & $2^{+1}_{-0}$ & $38\pm1$ & $2.6$ & $9.43\pm0.38$ \\
AEGIS-15569 & 14:19:50.977 & +52:53:25.728 & $1.674$ & $348\pm52$ & $8.40^{+0.17}_{-0.16}$ & $8^{+6}_{-3}$ & $31\pm4$ & $1.2$ & $8.90\pm0.40$ \\
AEGIS-19374 & 14:19:57.008 & +52:55:27.003 & $1.685$ & $380\pm198$ & $7.75^{+0.26}_{-0.29}$ & $32^{+40}_{-17}$ & $<51$ & $2.0$ & $<9.55$ \\
AEGIS-22858 & 14:19:55.093 & +52:55:55.815 & $1.397$ & $738\pm154$ & $7.56^{+0.21}_{-0.16}$ & $79^{+40}_{-34}$ & $<33$ & $1.0$ & $<8.89$ \\
AEGIS-26531 & 14:19:52.778 & +52:56:21.812 & $1.588$ & $303\pm31$ & $8.73^{+0.08}_{-0.08}$ & $9^{+2}_{-2}$ & $60\pm4$ & $1.2$ & $9.50\pm0.38$ \\
AEGIS-29378 & 14:19:47.585 & +52:56:07.873 & $1.683$ & $276\pm51$ & $8.55^{+0.11}_{-0.15}$ & $5^{+2}_{-1}$ & $26\pm5$ & $1.4$ & $8.85\pm0.43$ \\
AEGIS-34848 & 14:19:39.730 & +52:56:00.265 & $1.524$ & $200\pm28$ & $9.17^{+0.07}_{-0.08}$ & $5^{+4}_{-2}$ & $59\pm5$ & $1.2$ & $9.47\pm0.39$ \\
GOODS-N-13876 & 12:36:10.789 & +62:12:39.078 & $1.625$ & $581\pm85$ & $8.13^{+0.19}_{-0.25}$ & $39^{+45}_{-16}$ & $46\pm7$ & $1.6$ & $9.38\pm0.42$ \\
GOODS-N-18360 & 12:36:10.480 & +62:13:58.559 & $1.674$ & $923\pm32$ & $7.64^{+0.14}_{-0.12}$ & $81^{+33}_{-23}$ & $42\pm5$ & $1.0$ & $9.10\pm0.40$ \\
GOODS-N-18548 & 12:36:17.755 & +62:14:00.517 & $1.485$ & $337\pm16$ & $8.60^{+0.11}_{-0.15}$ & $5^{+3}_{-1}$ & $23\pm1$ & $2.8$ & $9.04\pm0.38$ \\
GOODS-N-19659 & 12:36:24.654 & +62:14:18.762 & $1.451$ & $289\pm 5$ & $8.84^{+0.09}_{-0.06}$ & $14^{+10}_{-4}$ & $59\pm1$ & $1.5$ & $9.57\pm0.38$ \\
GOODS-N-23634 & 12:36:27.007 & +62:15:29.858 & $1.676$ & $440\pm121$ & $8.31^{+0.16}_{-0.14}$ & $22^{+14}_{-9}$ & $<48$ & $2.5$ & $<9.60$ \\
GOODS-N-19149 & 12:36:32.669 & +62:14:11.360 & $1.383$ & $367\pm21$ & $8.18^{+0.12}_{-0.11}$ & $17^{+7}_{-4}$ & $45\pm2$ & $1.6$ & $9.38\pm0.38$ \\
GOODS-N-18817 & 12:36:40.516 & +62:14:03.574 & $1.485$ & $277\pm 6$ & $8.87^{+0.05}_{-0.05}$ & $10^{+2}_{-1}$ & $48\pm1$ & $2.0$ & $9.50\pm0.38$ \\
GOODS-N-26186 & 12:36:38.417 & +62:16:13.757 & $1.487$ & $707\pm264$ & $7.40^{+0.12}_{-0.08}$ & $165^{+38}_{-42}$ & $20\pm1$ & $1.6$ & $8.64\pm0.38$ \\
GOODS-N-22263 & 12:37:17.724 & +62:15:06.145 & $1.488$ & $501\pm24$ & $7.92^{+0.15}_{-0.16}$ & $22^{+15}_{-8}$ & $22\pm1$ & $1.2$ & $8.65\pm0.38$ \\
GOODS-N-25465 & 12:37:21.196 & +62:16:00.840 & $1.433$ & $166\pm15$ & $8.74^{+0.08}_{-0.09}$ & $8^{+2}_{-2}$ & $<24$ & $1.3$ & $<8.73$ \\
GOODS-N-29675 & 12:37:07.081 & +62:17:18.971 & $1.684$ & $561\pm43$ & $8.20^{+0.11}_{-0.12}$ & $41^{+18}_{-10}$ & $26\pm0$ & $1.4$ & $8.84\pm0.38$ \\
GOODS-N-29190 & 12:36:56.424 & +62:17:09.787 & $1.488$ & $219\pm 9$ & $8.59^{+0.07}_{-0.08}$ & $13^{+4}_{-3}$ & $49\pm2$ & $1.7$ & $9.47\pm0.38$ \\
GOODS-N-33726 & 12:36:59.343 & +62:18:52.358 & $1.450$ & $308\pm 9$ & $8.64^{+0.09}_{-0.08}$ & $6^{+3}_{-2}$ & $39\pm1$ & $1.6$ & $9.25\pm0.38$ \\
GOODS-N-33438 & 12:36:43.891 & +62:18:45.842 & $1.684$ & $207\pm18$ & $9.34^{+0.06}_{-0.07}$ & $9^{+3}_{-2}$ & $76\pm2$ & $2.4$ & $9.99\pm0.38$ \\
\hline
\end{tabular}
\caption{Coordinates, spectroscopic redshifts, [O~{\scriptsize III}]~$\lambda5007$ EWs, stellar masses, sSFRs, velocity dispersions, effective radii, and dynamical masses of the $34$ EELGs in our Sample I. Redshifts are derived from [O~{\scriptsize III}]~$\lambda5007$ emission lines. Stellar masses and sSFRs are derived from BEAGLE SED fitting with constant SFH models (Section~\ref{sec:beagle}). Velocity dispersions are computed from resolved [O~{\scriptsize III}]~$\lambda5007$ emission line widths, and effective radii are measured from {\it HST}/WFC3 F160W imaging. Dynamical masses are computed using velocity dispersions and effective radii (Section~\ref{sec:dyn_mass}).}
\label{tab:dyn_sample}
\end{table*}

%%%% Table: Properties of EELGs in Sample II %%%%

\begin{table*}
\begin{tabular}{|c|c|c|c|c|c|c|c|}
\hline
ID & R.A. & Decl. & $z_{\rm{spec}}$ & EW$_{\rm{[OIII]}\lambda5007}$ & $\log{(M_{\star,\rm{CSFH}}/M_{\odot})}$ & sSFR$_{\rm{CSFH}}$ & age$_{\rm{CSFH}}$ \\
 & (hh:mm:ss) & (dd:mm:ss) & & (\AA) & & (Gyr$^{-1}$) & (Myr) \\
\hline
AEGIS-04711 & 14:19:34.958 & $+$52:47:50.219 & $2.1839$ & $1060\pm 25$ & $2.3^{+0.1}_{-0.1}\times10^8$ & $118^{+9}_{-10}$ & $8.5^{+0.8}_{-0.7}$ \\
AEGIS-15778 & 14:19:11.210 & $+$52:46:23.414 & $2.1716$ & $1001\pm 42$ & $1.3^{+0.1}_{-0.1}\times10^8$ & $161^{+13}_{-13}$ & $6.2^{+0.6}_{-0.5}$ \\
UDS-08078 & 02:17:02.741 & $-$05:14:57.498 & $3.2277$ & $ 881\pm 20$ & $1.6^{+0.2}_{-0.1}\times10^9$ & $85^{+13}_{-14}$ & $11.7^{+2.4}_{-1.5}$ \\
UDS-09067 & 02:17:01.477 & $-$05:14:45.359 & $3.2288$ & $1694\pm 42$ & $3.0^{+0.3}_{-0.2}\times10^8$ & $136^{+12}_{-12}$ & $7.3^{+0.8}_{-0.6}$ \\
UDS-12539 & 02:17:53.733 & $-$05:14:03.196 & $1.6211$ & $ 882\pm 33$ & $1.1^{+0.1}_{-0.0}\times10^8$ & $188^{+8}_{-10}$ & $5.3^{+0.3}_{-0.2}$ \\
UDS-19167 & 02:17:43.535 & $-$05:12:43.610 & $2.1843$ & $1532\pm133$ & $7.2^{+0.2}_{-0.2}\times10^7$ & $237^{+11}_{-11}$ & $4.2^{+0.2}_{-0.2}$ \\
UDS-21724 & 02:17:20.006 & $-$05:12:10.624 & $3.2278$ & $1061\pm 34$ & $3.3^{+0.3}_{-0.3}\times10^8$ & $137^{+9}_{-10}$ & $7.3^{+0.7}_{-0.5}$ \\
\hline
\end{tabular}
\caption{Coordinates, spectroscopic redshifts, [O~{\scriptsize III}]~$\lambda5007$ EWs, stellar masses, sSFRs, and stellar ages of the $7$ the most extreme [O~{\scriptsize III}] emitters in our Sample II. The stellar masses, sSFRs, and stellar ages are derived from constant SFH models with BEAGLE (Section~\ref{sec:beagle}).}
\label{tab:irac_sample}
\end{table*}

%%%%%%%%%%%% SED FITTING %%%%%%%%%%%%

\section{Spectral energy distribution fitting} \label{sec:fitting}

We derive the physical properties (e.g., stellar mass) and constrain the stellar populations of EELGs in our Samples I and II from SED fitting. We first consider stellar population synthesis modeling with a constant star formation history using the BayEsian Analysis of GaLaxy sEds (BEAGLE, version 0.23.0; \citealt{Chevallard2016}) tool in Section~\ref{sec:beagle}. To better constrain potential older stellar populations ($>$ a few hundred Myr) in the most extreme [O~{\small III}] emitters in Sample II, we also perform SED fitting with nonparametric SFH models using the Bayesian Analysis of Galaxies for Physical Inference and Parameter EStimation (BAGPIPES; \citealt{Carnall2018}) in Section~\ref{sec:bagpipes}. 

\subsection{Constant SFH model fitting} \label{sec:beagle}

Following the procedures in \citetalias{Tang2019}, we model the broadband photometry and available emission line fluxes ([O~{\small II}], H$\beta$, [O~{\small III}], H$\alpha$) of the objects in Samples I and II using the BEAGLE tool. Here we use single stellar population models assuming a constant SFH (hereafter CSFH models). For the EELGs in Sample I, the stellar masses derived from CSFH model fitting are compared with dynamical masses in Section~\ref{sec:dyn_mass}. We also examine whether the CSFH models are able to recover the rest-frame NIR luminosities of the most extreme [O~{\small III}] emitters in Sample II, which may probe the hidden older stellar populations that might be masked by very young stars ($<10$~Myr) at rest-frame UV-to-optical wavelengths. 

Details of the BEAGLE modeling have been described in \citetalias{Tang2019} and we briefly summarise in the following. BEAGLE adopts the combination of the latest version of the \citet{Bruzual2003} stellar population synthesis models and the photoionisation models of star-forming galaxies of \citet{Gutkin2016} with {\small CLOUDY} \citep{Ferland2013}. We adopt a \citet{Chabrier2003} initial mass function (IMF) and allow the metallicity to vary in the range $-2.2 \le \log{(Z/Z_{\odot})} \le 0.25$ ($Z_{\odot}=0.01524$; \citealt{Caffau2011}). The gas-phase metallicity is set to equal to the stellar metallicity. The electron density is fixed to $n_e=100$~cm$^{-3}$ consistent with the density inferred from typical star-forming galaxies at $z\sim2$ \citep[e.g.,][]{Sanders2016,Steidel2016}. The ionisation parameter $U$ and the dust-to-metal ratio $\xi_{\rm{d}}$ are adjusted in the range $-4.0\le \log{U_{\rm{S}}}\le-1.0$ and $0.1\le\xi_{\rm{d}}\le0.5$. We assume the \citet{Calzetti2000} extinction curve to account for the dust attenuation in the neutral interstellar medium (ISM), and we adopt the prescription of \citet{Inoue2014} to include the absorption of intergalactic medium (IGM).

The best-fitting stellar masses and sSFRs are presented in Table~\ref{tab:dyn_sample}. We find similar stellar mass and sSFR versus [O~{\small III}]~$\lambda5007$ EW trends for Sample I as in \citetalias{Tang2019}, namely that galaxies with the largest [O~{\small III}]~$\lambda5007$ EWs ($>800$~\AA) have the lowest stellar masses ($M_{\star}\simeq10^7-10^8\ M_{\odot}$) and undergo intense bursts of star formation (sSFR $\gtrsim100$~Gyr$^{-1}$). For objects in Sample II, we fit the rest-frame UV-to-NIR SEDs with CSFH models as their robust IRAC (rest-frame NIR) fluxes are available, and the best-fitting stellar masses are presented in Table~\ref{tab:irac_sample}.

\subsection{Nonparametric SFH model fitting} \label{sec:bagpipes}

Nonparametric SFH fitting has the advantage it can recover more complex SFHs of galaxies \citep[e.g.,][]{Tojeiro2007,Pacifici2016,Iyer2019,Leja2019,Lower2020,Tacchella2022a}. In order to better reconstruct the potential past SFHs of the most extreme [O~{\small III}] emitting galaxies, we use nonparametric SFH stellar population models to fit the rest-frame UV-to-NIR SEDs of the $7$ objects in Sample II using BAGPIPES. BAGPIPES uses the 2016 version of the \citet{Bruzual2003} stellar population synthesis models with a \citet{Kroupa2001} IMF, and implements nebular emission models constructed using the {\small CLOUDY} photoionisation code following the methodology of \citet{Byler2017}. We allow the metallicity to vary from $0$ to $2.5\ Z_{\odot}$. The ionisation parameter is fixed to $\log{U}=-2.0$, which is consistent with the typical ionisation parameter derived for the most extreme line emitters from BEAGLE \citep{Tang2021a,Tang2021b}. We assume the \citet{Calzetti2000} extinction curve, with the dust attenuation ($A_V$) varies in the range $0-2$. 

In order to recover the presence of earlier stellar populations, we fit the observed SEDs with nonparametric models for the mass formed in a series of piecewise constant functions in lookback time. With BAGPIPES we adopt the following seven time bins in models (where $t$ represents the lookback time):
\begin{eqnarray*}
0<&t&<3\ \rm{Myr}; \\
3<&t&<10\ \rm{Myr}; \\
10<&t&<30\ \rm{Myr}; \\
30<&t&<100\ \rm{Myr}; \\
100<&t&<300\ \rm{Myr}; \\
300\ \rm{Myr}<&t&<1\ \rm{Gyr}; \\
1\ \rm{Gyr}<&t&<4\ \rm{Gyr}.
\end{eqnarray*}
Each time bin is spaced equally in logarithmic scale except the first and the last bin, as is common practice in the use of nonparametric SFH studies and it is more scalable in a sampling framework \citep[e.g.,][]{Leja2017,Leja2019,Tacchella2022a}. Such an approach is also consistent with \citet{Ocvirk2006} who find that the distinguish ability of simple stellar populations is roughly proportional to their separation in logarithmic time. For each time bin, we assume a constant SFH and fit the stellar mass formed in the bin as a free parameter (in the range $1<\log{(M_{\star}/M_{\odot})}<15$; the $\log{M}$ prior, see \citealt{Leja2019}). The BAGPIPES SED fitting is performed using Bayesian statistical techniques with nested sampling algorithms. The code outputs the posterior distribution of the stellar mass formed in each time bin and we compute the corresponding star formation rate. We will describe the stellar masses and stellar populations of the most extreme [O~{\small III}] emitters in Sample II derived from both parametric and nonparametric model fitting in Section~\ref{sec:stellar_mass}.

%%%%%%%%%%%% RESULTS %%%%%%%%%%%%

\section{Constraining evolved stellar populations in the most extreme [O~{\small III}] emitters} \label{sec:old_star}

In this section, we address the possibility of evolved stellar populations in the most extreme [O~{\small III}] emitting galaxies using dynamical mass measurements and SFHs derived from SED fitting. We first quantify the dependence of the dynamical mass and the dynamical-to-stellar mass ratio on [O~{\small III}] EW for the objects in our Sample I (Section~\ref{sec:dyn_mass}). We then characterize the stellar populations and SFHs of the most extreme [O~{\small III}] emitting galaxies by fitting the rest-frame UV-to-NIR SEDs of the objects in Sample II (Section~\ref{sec:stellar_mass}). 

\subsection{Dynamical masses of extreme [O~{\small III}] emitters} \label{sec:dyn_mass}

The most intense optical line emitting galaxies have been found to have very young stellar ages ($<10$~Myr) and low stellar masses by fitting SEDs with constant SFH stellar population models \citepalias{Tang2019}. If there are hidden older stellar populations in these systems, we would expect very large dynamical masses compared to the stellar masses inferred from CSFH models, and hence an increasing dynamical-to-stellar (CSFH) mass ratio with [O~{\small III}] EW or sSFR (derived from CSFH models). The dynamical masses are computed using velocity dispersions measured from resolved [O~{\small III}]~$\lambda5007$ emission line and half-light radii, and adopt the equation in \citet{Maseda2013}:
\begin{eqnarray}
M_{\rm{dyn}}=C\frac{r_{\rm{eff}}\sigma^2}{G},
\end{eqnarray}
where $\sigma$ is the velocity dispersion and $r_{\rm{eff}}$ is the half-light radius. The typical uncertainty of the half-light radius of EELGs at $z\sim2$ is $10$~per~cent \citep{vanderWel2012,Maseda2014}, and we adopt this in estimating the uncertainty of dynamical mass. The factor $C$ depends on the kinematic properties of galaxies. According to \citet{Price2016}, dispersion-dominated galaxies result in $C\approx5$, while $C\approx2.7$ is adopted for rotation-dominated galaxies. \citet{Erb2006} assume a disk geometry and derive $C\approx3.4$. In order to be consistent with other studies of emission line galaxies at $z\sim1-2$ \citep[e.g.,][]{Maseda2014,Masters2014}, we adopt $C=3$ as used in \citet{Maseda2013} with a conservative uncertainty of $33$~per~cent \citep[e.g.,][]{Rix1997}. 

\begin{figure*}
\begin{center}
\includegraphics[width=\linewidth]{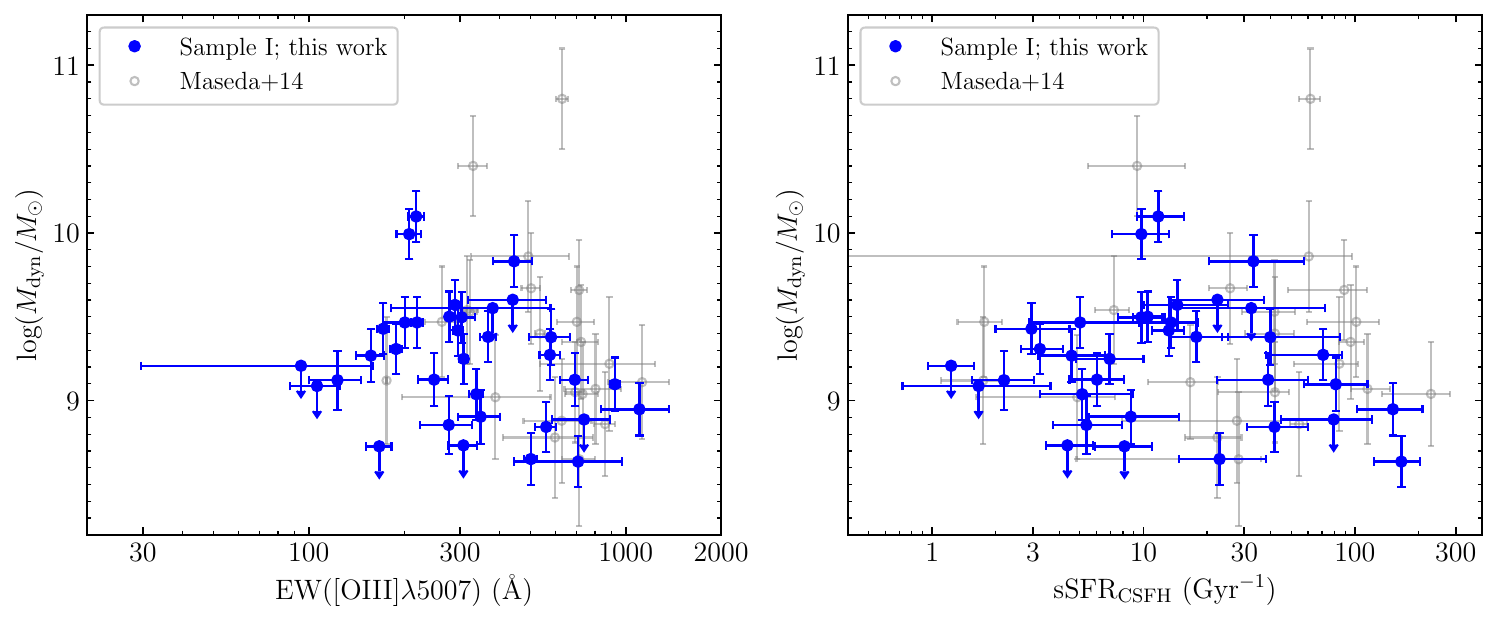}
\caption{The dynamical mass versus [O~{\scriptsize III}]~$\lambda5007$ EW (left panel) and sSFR (right panel) for the $34$ EELGs in our Sample I (blue solid circles) and the \citet{Maseda2014} EELGs at $z\sim1-2$ (grey open circles). Objects with velocity dispersion smaller than instrument resolution are shown as $3\sigma$ upper limits. Galaxies with larger [O~{\scriptsize III}] EWs or sSFRs tend to have lower dynamical masses, though with large scatter.}
\label{fig:dyn_mass}
\end{center}
\end{figure*}

The dynamical masses of the EELGs in Sample I are presented in Table~\ref{tab:dyn_sample}; they range from $10^{8.6}\ M_{\odot}$ to $10^{10.1}\ M_{\odot}$ with a median value of $10^{9.3}\ M_{\odot}$. These are systematically lower than the dynamical masses of typical $z\sim2$ star-forming galaxies ($\sim10^{10}-10^{11}\ M_{\odot}$; e.g., \citealt{Erb2006,Price2016}). In Fig.~\ref{fig:dyn_mass}, we show the dynamical mass as functions of the [O~{\small III}]~$\lambda5007$ EW and sSFR (derived from CSFH models) for our sample, and also the dynamical masses of the $22$ EELGs at $z\sim1-2$ in \citet{Maseda2014}. We notice that the \citet{Maseda2014} sample has slightly larger dynamical masses at fixed [O~{\small III}] EW or sSFR compared to our sample as a result of their brighter targets. For both samples we find a moderate correlation between dynamical mass and [O~{\small III}] EW (Spearman correlation coefficient $\rho=-0.45$ and $p$-value $p=1.9\times10^{-2}$) and a weak correlation between dynamical mass and sSFR ($\rho=-0.18$, $p=0.37$), that galaxies with larger [O~{\small III}] EWs or sSFRs have lower dynamical masses. For the most extreme line emitters with [O~{\small III}]~$\lambda5007$ EW $>800$~\AA\ or sSFR$_{\rm{CSFH}}>100$~Gyr$^{-1}$, the dynamical masses (median $M_{\rm{dyn}}=10^{9.1}\ M_{\odot}$) are $\sim2\times$ lower than those of galaxies with lower EWs (median $M_{\rm{dyn}}=10^{9.3}\ M_{\odot}$). This confirms the previous findings that the most extreme optical line emitting galaxies are low-mass systems (e.g., \citealt{Reddy2018}; \citetalias{Tang2019}; \citealt{Sanders2020}). 

\begin{figure*}
\begin{center}
\includegraphics[width=\linewidth]{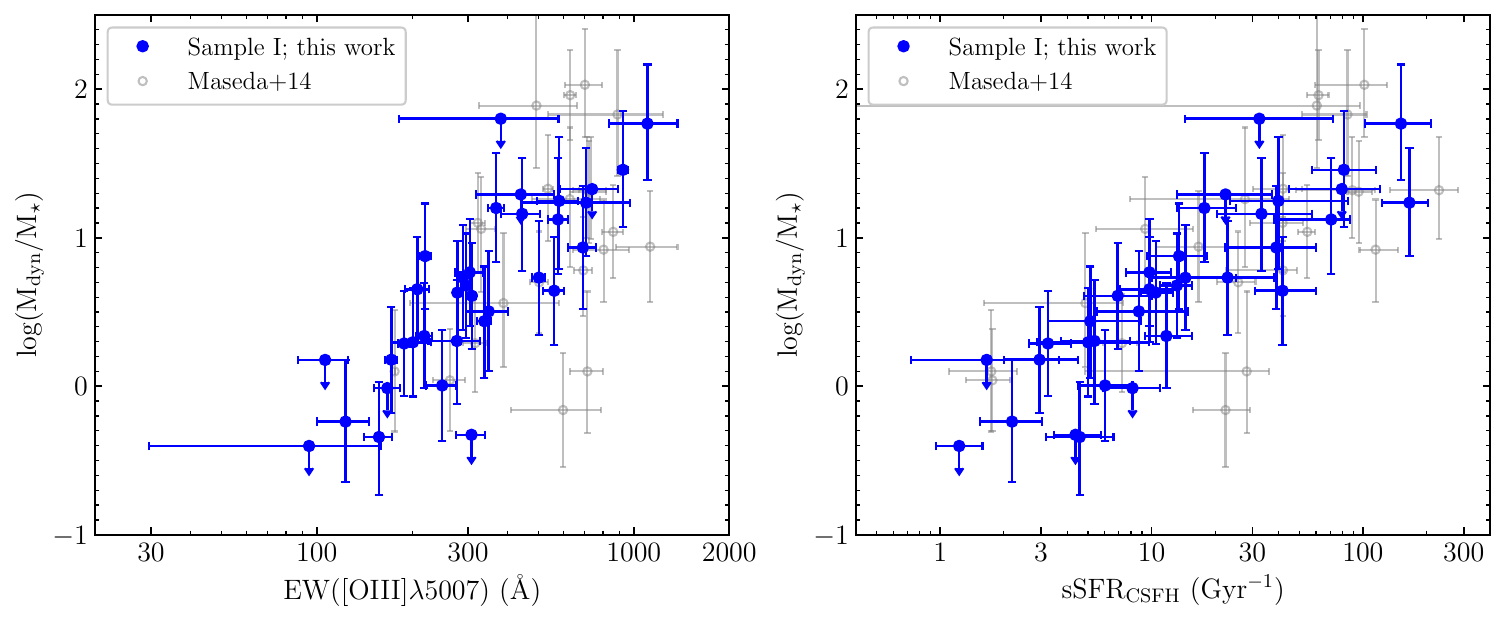}
\caption{The dynamical-to-stellar mass ratio (assuming single stellar population models with CSFH) versus [O~{\scriptsize III}]~$\lambda5007$ EW (left panel) and sSFR (inferred from single population CSFH models; right panel) for the $34$ EELGs in our Sample I (blue solid circles) and the \citet{Maseda2014} EELGs at $z\sim1-2$ (grey open circles). A clear trend is shown that galaxies with larger [O~{\scriptsize III}] EWs or sSFRs have larger dynamical-to-stellar mass ratios.}
\label{fig:dyn_star_ratio}
\end{center}
\end{figure*}

We next constrain the presence of older stellar populations in the most extreme [O~{\small III}] emitters by comparing the dynamical mass to the stellar mass inferred from CSFH models. In Fig.~\ref{fig:dyn_star_ratio}, we plot the dynamical-to-stellar mass ratios of the objects in Sample I (blue solid circles) together with those of the \citet{Maseda2014} sample (grey open circles). In order to be consistent, we re-compute the stellar masses of the EELGs in \citet{Maseda2014} with BEAGLE, assuming single stellar population models with CSFH and following the same procedures as for our objects (see Section~\ref{sec:beagle}). It is remarkable that the dynamical-to-stellar mass ratio is strongly correlated with [O~{\small III}] EW (Spearman correlation coefficient $\rho=0.84$ and $p$-value $p=3.5\times10^{-8}$) and sSFR ($\rho=0.90$, $p=1.1\times10^{-10}$), that the ratio increases with [O~{\small III}] EW and sSFR$_{\rm{CSFH}}$ for both EELG samples. The median dynamical-to-stellar mass ratio of galaxies with [O~{\small III}]~$\lambda5007$ EW $<300$~\AA\ is $M_{\rm{dyn}}/M_{\star,\rm{CSFH}}=2$, and then this value increases to $10$ for galaxies with [O~{\small III}]~$\lambda5007$ EW $=300-800$~\AA\ and sSFR$_{\rm{CSFH}}\simeq10$~Gyr$^{-1}$ (i.e., the average [O~{\small III}] EW and sSFR of typical $z\simeq7-8$ star-forming galaxies; e.g., \citealt{Labbe2013,Endsley2021}). For galaxies with the largest [O~{\small III}]~$\lambda5007$ EWs ($>800$~\AA) and sSFRs ($>100$~Gyr$^{-1}$), the median dynamical-to-stellar mass ratio is $M_{\rm{dyn}}/M_{\star,\rm{CSFH}}\simeq20$ with a maximum reaching $M_{\rm{dyn}}/M_{\star,\rm{CSFH}}\simeq100$. Previous studies of more massive star-forming galaxies at $z\sim2$ have also shown a positive correlation between the dynamical-to-stellar mass ratio and sSFR \citep[e.g.,][]{Price2016} or H$\alpha$ EW \citep[e.g.,][]{Erb2006}. The increase of dynamical-to-stellar mass ratio with optical line EW and sSFR$_{\rm{CSFH}}$ indicates that the mass of recently formed stars ($<10$~Myr assuming CSFH) in the most intense line emitting galaxies comprises only $\sim1-10$~per~cent of the total dynamical mass. This suggests the dominant mass must arise from other components such as dark matter, gas, and perhaps the older stellar populations. We investigate each possibility in turn.

Regarding dark matter, recent studies \citep[e.g.,][]{Wuyts2016,Price2020} have compared the baryonic mass (i.e., stellar mass and gas mass) to the dynamical mass for typical $z\sim2$ star-forming galaxies. The results show that dark matter contributes only a small fraction ($\lesssim10$~per~cent) to the total dynamical mass. Assuming these results are representative for our sample, it suggests the bulk of the excess mass must be baryonic (i.e., gas or evolved stellar populations).

As our EELGs are undergoing intense bursts of star formation, it is likely that these systems have a large gas fraction. Ignoring for the moment a contribution from from evolved stars, we infer that the gas fraction must approach $\sim80-90$~per~cent of the dynamical mass (assuming a dark matter fraction of $10$~per~cent). The commonly-used Kennicutt-Schmidt (KS) law \citep{Kennicutt1998} is likely inapplicable here since starburst galaxies have higher star formation efficiencies  \citep[e.g.,][]{Bouche2007,Genzel2010,Wuyts2016}. Using gas masses derived from CO or far-infrared emission, recent studies find that the gas fraction increases with sSFR \citep[e.g.,][]{Dessauges-Zavadsky2015,Genzel2015,Schinnerer2016}, reaching to $\sim50-90$~per~cent at sSFR $\sim10$~Gyr$^{-1}$. This is lower or only marginally comparable to that required to explain our dynamical-to-stellar mass ratios in the absence of older stars. \citet{Maseda2014} also derive gas fractions for their sample of $z\sim1-2$ EELGs based on the Jeans and Toomre instability criteria and quote values of $>67$~per~cent. In summary, it is still unclear whether our large dynamical-to-stellar mass ratios can be explained solely due to gaseous reservoirs. In the next subsection, we will provide new constraints on the presence of older stars by fitting the rest-frame UV-to-NIR SEDs.

%%%% Figure: Dynamical mass vs. [OIII] EW & sSFR %%%%

%%%% Figure: Dynamical-to-stellar mass ratio vs. [OIII] EW & sSFR %%%%

\subsection{Stellar populations and star formation histories of the most extreme [O~{\small III}] emitting galaxies} \label{sec:stellar_mass}

The final possibility for the large dynamical-to-stellar mass ratios is the presence of much older ($>$ a few hundred Myr) stars whose rest-frame UV-to-optical light is obscured by a young starburst. Such older stellar populations could be revealed via the SEDs of galaxies at rest-frame NIR wavelengths. In order to constrain the contribution of old stellar populations in the most extreme line emitters, we derive the stellar masses and SFHs of the $7$ galaxies with [O~{\small III}]~$\lambda5007$ EW $>800$~\AA\ and robust IRAC detections in our Sample II by fitting their rest-frame UV-to-NIR SEDs. 

\begin{figure*}
\begin{center}
\includegraphics[width=\linewidth]{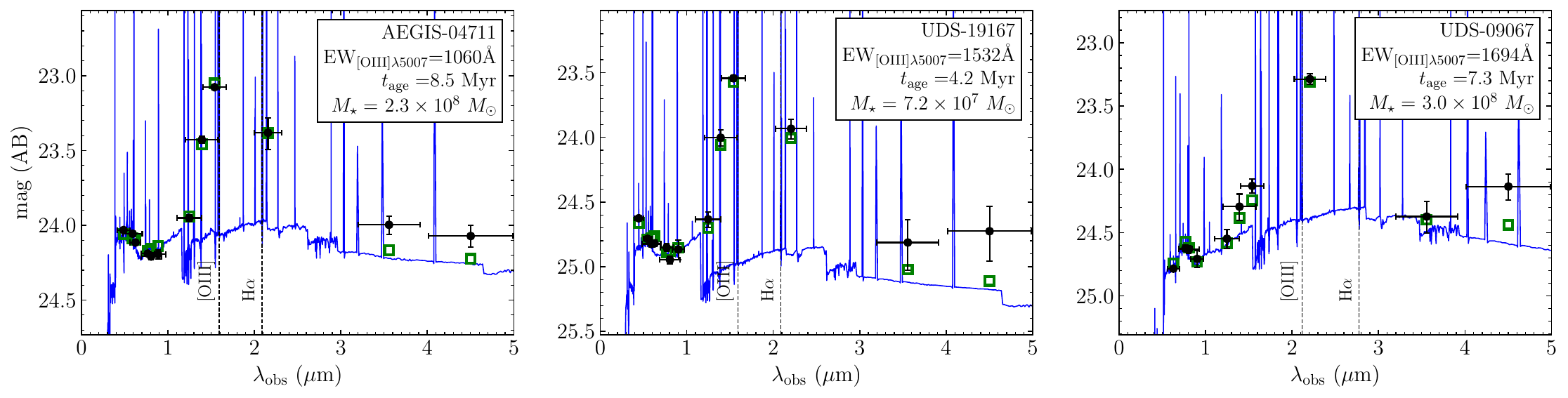}
\caption{Broadband SEDs and the best-fitting CSFH models (derived from BEAGLE) of the most extreme [O~{\scriptsize III}] emitters at $z=1.3-3.7$ with robust IRAC flux measurements in Sample II. Observed broadband photometry is shown as solid black circles. The best-fitting BEAGLE SED models are plotted by solid blue lines, and synthetic photometry is shown as open green squares. Strong rest-frame optical emission lines, [O~{\scriptsize III}]$\lambda5007$ and H$\alpha$, are highlighted by dashed black lines. Although the rest-frame UV-to-optical SEDs can be well reproduced by CSFH models, the rest-frame NIR (IRAC) luminosities for these objects are underestimated.}
\label{fig:irac_sed}
\end{center}
\end{figure*}

We first fit SEDs of the $7$ objects in Sample II using the constant CSFH models introduced in Section~\ref{sec:beagle} with BEAGLE. The goal of this step is to investigate whether a single component stellar population model is able to reproduce the full observed SEDs especially at rest-frame NIR wavelengths. By fitting SEDs with CSFH models, we derive best-fitting stellar ages of the $7$ galaxies in Sample II ranging from $4$ to $12$~Myr. The stellar masses of these young systems are from $7\times10^7\ M_{\odot}$ to $1.6\times10^9\ M_{\odot}$ (Table~\ref{tab:irac_sample}). Although CSFH models can reproduce the rest-frame UV-to-optical SEDs of the most extreme line emitting sources, such models reproduce the observed IRAC (i.e., rest-frame NIR) luminosities for only $2$ of the $7$ objects in Sample II (UDS-08078, UDS-21724); they underestimate the IRAC luminosities for $5$ objects (AEGIS-04711, AEGIS-15778, UDS-09067, UDS-12539, UDS-19167). In Fig.~\ref{fig:irac_sed}, we plot rest-frame UV-to-NIR SEDs and the best-fitting CSFH models for the objects in Sample II. As shown in the figure, CSFH models only reproduce $50-80$~per~cent of the observed IRAC luminosities, well below the observed $1\sigma$ lower limit.

We next fit SEDs of the objects in Sample II using nonparametric SFH models with BAGPIPES. As demonstrated in Section~\ref{sec:bagpipes}, we will derive the stellar masses formed in the seven lookback time bins from the most recent $3$~Myr to $>1$~Gyr ago. We aim to constrain the presence of possible older stellar populations in the most extreme optical line emitters, and whether the rest-frame NIR luminosities can be reproduced by including such stars. Note that in the following we will exclude UDS-21724 from our nonparametric SFH modeling since the strong nebular emission of this object cannot be well fitted by BAGPIPES which will result in an overestimation of the stellar mass.

\begin{figure*}
\begin{center}
\includegraphics[width=0.95\linewidth]{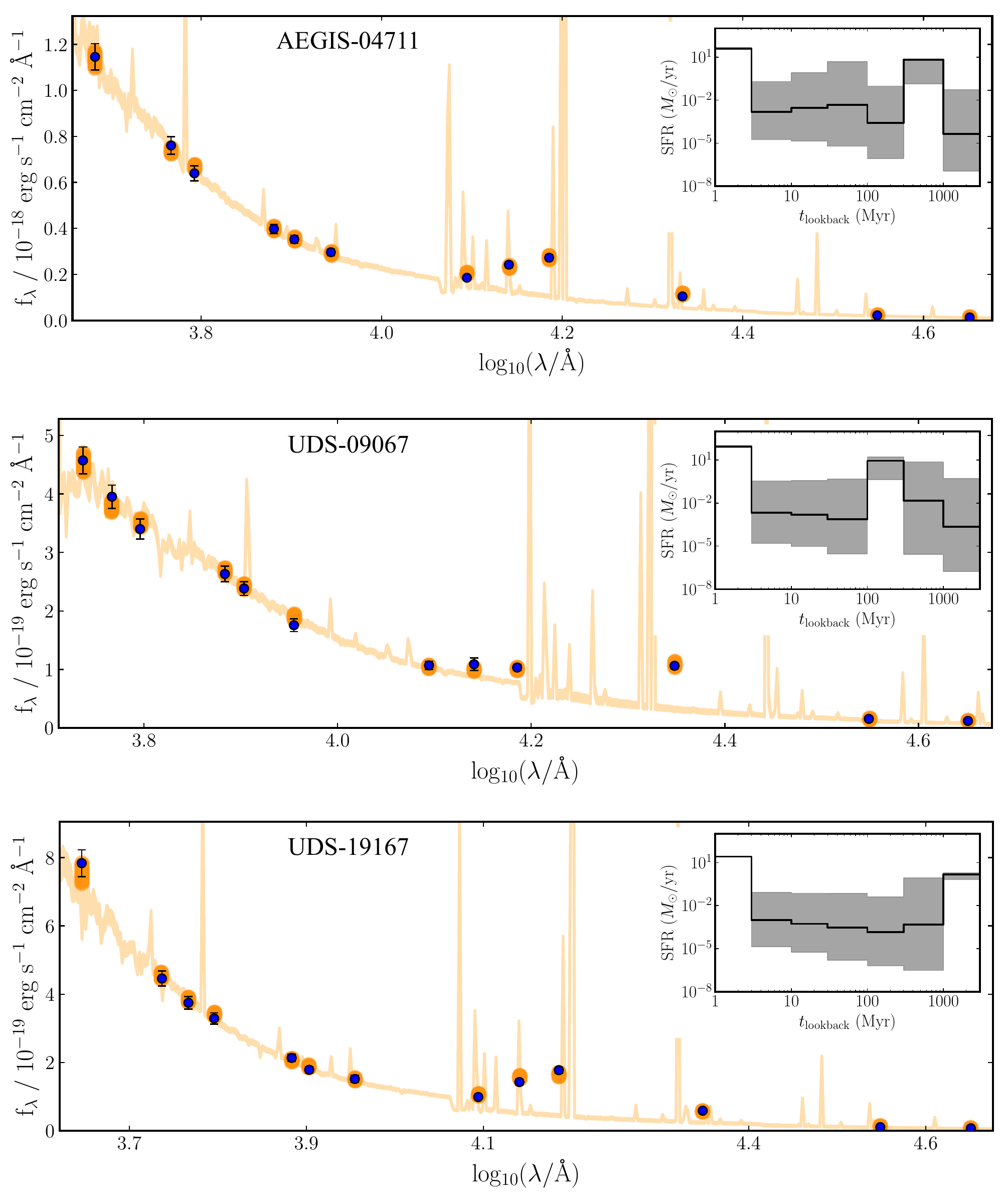}
\caption{Broadband SEDs and the best-fitting nonparametric SFH models (derived from BAGPIPES) of the most extreme [O~{\scriptsize III}] emitting galaxies in Sample II. Observed broadband fluxes including uncertainties are shown by blue circles, and the best-fitting model and the synthetic photometry (with the $84$th and $16$th percentiles) are shown by the orange line and the orange points. In the upper right of each SED plot, we show the SFH recovery of each object. The SFH is given by SFRs derived in the seven lookback time bins described in Section~\ref{sec:bagpipes}. Posterior median SFRs are plotted by black lines, and the grey shaded regions show the $16$th to $84$th percentiles (i.e., the $\pm1\sigma$ uncertainties). The nonparametric SFH models can well reproduce the SEDs including the IRAC (rest-frame NIR) luminosities, and reconstruct the past SFHs ($>100$~Myr) for the most extreme [O~{\scriptsize III}] emitters in addition to the intense bursts of star formation within $<10$~Myr.}
\label{fig:npsfh}
\end{center}
\end{figure*}

The best-fitting BAGPIPES nonparametric SFH models for Sample II are plotted in Fig.~\ref{fig:npsfh}. In contrast to the CSFH models, the SEDs and IRAC luminosities can be well reproduced within $1\sigma$ uncertainty by nonparametric SFH models. In Table~\ref{tab:npsfh_irac} we present the stellar masses formed in the seven time bins for the objects in Sample II. We notice that the stellar masses formed in the first $10$~Myr inferred from nonparametric SFH models are from $5\times10^7\ M_{\odot}$ to $5\times10^8\ M_{\odot}$, roughly consistent with the stellar masses derived from CSFH models. More remarkably, however, a significant fraction of the total stellar mass was formed at $>100$~Myr ago. The evolved stellar masses of galaxies in Sample II range from $3\times10^9\ M_{\odot}$ to $1\times10^{10}\ M_{\odot}$, i.e. much greater than those associated with the secondary burst phase ($<10$~Myr). The results suggest that the rest-frame NIR light of these systems is likely dominated by stellar populations formed over a few hundred Myr ago, which cannot be easily identified at rest-frame UV-to-optical wavelengths. 

The recovered SFHs from nonparametric models for objects in Sample II are also shown in Fig.~\ref{fig:npsfh}. We notice that the models predict a ``two-burst''-like SFH, the most recent $\le3$~Myr earlier following a first event between $100$~Myr and $1$~Gyr earlier. When EELGs are in the current burst phase, the massive stars are likely being formed in very young star clusters as demonstrated by spatially-resolved observations of a few strongly lensed galaxies at high redshift \citep[e.g.][]{Vanzella2019,Vanzella2022}. When not in their present burst phase, they were forming stars with a negligible rate (lower than $10^{-2}\ M_{\odot}$~yr$^{-1}$ with large uncertainties). Such a low SFR implies a UV magnitude fainter than $31$ AB mag at $z\sim2$, below the detection limit of current {\it HST} and even upcoming {\it JWST} imaging surveys \citep{Robertson2021}. However, we note that the recovered SFH does not necessarily mean these systems are actually in the quiescent phase between the two ``bursts''. As reflected by the large uncertainties of SFRs (Fig.~\ref{fig:npsfh}) and stellar masses (Table~\ref{tab:npsfh_irac}) formed between $\sim10$~Myr and a few hundred Myr earlier, it is possible that the objects in Sample II followed a more gradual evolution during this period. The key point is that starlight from this period could be outshone by young massive stars at rest-frame UV-to-optical and by older stars at rest-frame NIR wavelengths. Nevertheless, we emphasise that the ``burst'' phase happened $>100$~Myr ago reflects the presence of evolved stellar populations in these systems.

Finally, we examine whether the extremely large dynamical-to-stellar mass ratios found in Section~\ref{sec:dyn_mass} could be explained by introducing evolved stellar populations inferred from nonparametric SFH fitting. Compared to the young stellar masses formed in the first $10$~Myr, the stellar masses formed at $>100$~Myr are $7-100\times$ (with a median of $39\times$) larger (Table~\ref{tab:npsfh_irac}), amounting to $87-99$~per~cent (with a median of $97$~per~cent) of the total stellar mass. This is consistent with studies of local extreme [O~{\small III}] line emitting ``Green Pea'' galaxies \citep{Cardamone2009} where only $\sim4-20$~per~cent of their stellar masses are produced in the most recent burst \citep{Amorin2012}. If the most intense line emitters in Sample II follow the $M_{\rm{dyn}}/M_{\star,\rm{CSFH}}$ - EW relation derived from our Sample I, the result could explain the large dynamical-to-stellar mass (derived from CSFH models) ratios ($M_{\rm{dyn}}/M_{\star,\rm{CSFH}}\simeq10-100$) found for galaxies with [O~{\small III}]~$\lambda5007$ EW $>800$~\AA. The dynamical mass reflects not only the total stellar mass, but also the gas mass within the effective radius. Assuming the median old-to-young stellar mass ratio $\simeq39$ derived from our Sample II, and the dynamical-to-stellar mass ratio $M_{\rm{dyn}}/M_{\star,\rm{CSFH}}\simeq10-100$ found for the most extreme line emitters in Sample I, the gas fraction ($f_{\rm{gas}}=M_{\rm{gas}}/M_{\rm{dyn}}$) of the most extreme line emitters would be $60$~per~cent or less\footnote{Here we neglect the mass of dark matter within the effective radius since it only contributes a small fraction ($<10$~per~cent) to the dynamical mass \citep[e.g.][]{Wuyts2016,Price2020}}. This is somewhat lower than the gas fraction derived for EELGs at $z\sim1-2$ ($f_{\rm{gas}}\simeq2/3$) in \citet{Maseda2014}. On the other hand, if we assume the $f_{\rm{gas}}\simeq2/3$ in \citet{Maseda2014}, the evolved stellar mass needs to be $\simeq3-33\times$ the young stellar mass in order to explain the dynamical-to-stellar mass ratio at EW$_{\rm{[OIII]}}>800$~\AA\ in Sample I, which is lower than the values derived in our Sample II. However, we consider this may be due to the following reasons. First, the current size of Sample II is small, and we focus on the subset with robust rest-frame NIR photometry detections which might bias the sample towards systems with larger evolved stellar mass (and hence brighter rest-frame NIR luminosity). Second, the gas fraction or the old-to-young stellar mass ratio may vary with [O~{\small III}] EW, and the objects in our Sample II have larger [O~{\small III}] EWs comparing to the average EW of the sample in \citet{Maseda2014}. To test this scenario we need to compare with the gas fraction of the EW$_{\rm{[OIII]}}>800$~\AA\ galaxies in \citet{Maseda2014}. However, there are only a handful (three) of such objects so currently the statistics are not good enough to make such comparison. Given the fact that the rest-frame UV-to-optical luminosities of the most intense optical line emitting galaxies are dominated by very young stellar populations, the SED fitting results demonstrate that the rest-frame NIR luminosity provides a valuable probe of the evolved stellar populations in these systems as reflected by their dynamical masses. In Section~\ref{sec:discussion}, we discuss the implications for the similar sources in the reionisation era.

%%%% Figure: SED fitting with IRAC photometry %%%%

%%%% Figure: Nonparametric SFH fitting %%%%

%%%% Table: SFHs inferred from nonparametric SFH fitting with IRAC fluxes %%%%

\begin{table*}
\begin{tabular}{|c|c|c|c|c|c|c|c|}
\hline
Target ID & $\log{(M_{\star}/M_{\odot})}$ & $\log{(M_{\star}/M_{\odot})}$ & $\log{(M_{\star}/M_{\odot})}$ & $\log{(M_{\star}/M_{\odot})}$ & $\log{(M_{\star}/M_{\odot})}$ & $\log{(M_{\star}/M_{\odot})}$ & $\log{(M_{\star}/M_{\odot})}$ \\
 & $0-3$~Myr & $3-10$~Myr & $10-30$~Myr & $30-100$~Myr & $100-300$~Myr & $300$~Myr $-1$~Gyr & $>1$~Gyr \\
\hline
AEGIS-04711 & $8.07^{+0.05}_{-0.05}$ & $4.03^{+2.11}_{-1.92}$ & $4.75^{+2.46}_{-2.31}$ & $5.51^{+3.02}_{-2.88}$ & $4.71^{+2.58}_{-2.47}$ & $9.65^{+0.05}_{-1.68}$ & $4.94^{+3.08}_{-2.58}$ \\
AEGIS-15778 & $7.67^{+0.16}_{-4.10}$ & $6.17^{+1.79}_{-3.21}$ & $4.30^{+2.32}_{-2.22}$ & $4.59^{+2.38}_{-2.30}$ & $4.92^{+3.06}_{-2.67}$ & $9.67^{+0.05}_{-0.07}$ & $5.34^{+2.66}_{-2.86}$ \\
UDS-08078 & $8.67^{+0.07}_{-2.94}$ & $5.43^{+3.41}_{-2.95}$ & $4.50^{+2.53}_{-2.34}$ & $5.00^{+2.61}_{-2.64}$ & $9.87^{+0.08}_{-4.57}$ & $6.84^{+3.25}_{-3.85}$ & $5.33^{+3.01}_{-2.82}$ \\
UDS-09067 & $8.41^{+0.05}_{-0.05}$ & $4.17^{+2.20}_{-2.11}$ & $4.49^{+2.39}_{-2.22}$ & $4.73^{+2.79}_{-2.43}$ & $9.25^{+0.27}_{-1.32}$ & $7.01^{+2.68}_{-3.74}$ & $5.65^{+3.37}_{-3.12}$ \\
UDS-12539 & $7.89^{+0.07}_{-0.12}$ & $7.36^{+0.33}_{-0.46}$ & $4.23^{+2.15}_{-1.98}$ & $4.27^{+2.16}_{-2.13}$ & $3.85^{+2.36}_{-1.81}$ & $4.49^{+2.15}_{-2.06}$ & $10.02^{+0.03}_{-0.03}$ \\
UDS-19167 & $7.89^{+0.04}_{-0.03}$ & $3.85^{+1.92}_{-1.87}$ & $4.03^{+2.11}_{-1.98}$ & $4.33^{+2.37}_{-2.29}$ & $4.44^{+2.47}_{-2.32}$ & $5.52^{+3.25}_{-3.19}$ & $9.49^{+0.11}_{-0.38}$ \\
\hline
\end{tabular}
\caption{Stellar masses formed in the seven lookback time bins for the most extreme [O~{\small III}] emitting galaxies in Sample II. The time bins are introduced in Section~\ref{sec:bagpipes}, and the results are derived from BAGPIPES nonparametric SFH model fitting. A significant fraction of the total stellar mass is from evolved star formed at $>100$~Myr ago, and the mass of very young stellar populations ($<10$~Myr) compose a subdominant fraction of the total stellar mass.}
\label{tab:npsfh_irac}
\end{table*}

%%%%%%%%%%%% DISCUSSION %%%%%%%%%%%%

\section{Implications for stellar populations of galaxies in the reionisation era} \label{sec:discussion}

The results described in Section~\ref{sec:old_star} have suggested the possible presence of a significant population of evolved stars (age $>100$~Myr) in the most intense [O~{\small III}] emitters at $z=1.3-3.7$. The evidence is based on both the extremely large dynamical masses compared to that derived for the young ($<10$~Myr) stellar population, and nonparametric SFHs recovered from fitting the rest-frame UV-to-NIR photometry. Although galaxies with EW$_{\rm{[OIII]\lambda5007}}>800$~\AA\ are very rare at intermediate redshift \citep{Boyett2021}, this population is common in the reionisation era, comprising $20$~per~cent at $z\sim7$ \citep{Endsley2021}. Assuming our $z\sim1-3$ EELG sample are representative of the sources at higher redshift, we consider the implications of our results for line emitting galaxies in the reionisation era.

Our results suggest that the stellar light associated with an evolved population would be masked by both the stellar and nebular emission at rest-frame UV and optical wavelengths from young starbursts. Upcoming {\it JWST} surveys with the Near Infrared Camera (NIRCam; \citealt{Rieke2005}) will target the rest-frame UV-to-optical imaging for a large population of galaxies at $z\gtrsim7$, enabling more robust derivations of their stellar masses, SFRs and stellar ages \citep[e.g.][]{Tacchella2022b}. Meanwhile, the analyses of our $z\simeq1-3$ analogues suggest that the stellar masses and ages of the $z\simeq7-8$ galaxies with the most extreme [O~{\small III}] line emission may be significantly underestimated if they are based solely on analysing the rest-frame UV-to-optical photometry.

To illustrate this, we generate a mock galaxy spectrum at $z=8$ by adding a burst population (age $=5$~Myr) superposed on an evolved population with age $=300$~Myr following an instantaneous burst. Such a two-component system is consistent with a galaxy that first formed at $z\simeq12$ and underwent a secondary burst phase of star formation at $z=8$. We use the latest \citet{Bruzual2003} stellar population synthesis models and incorporate nebular emission computed from the {\small CLOUDY} code. We assume a sub-solar metallicity ($Z=0.2Z_{\odot}$) and an ionisation parameter $\log{U}=-2$, consistent with recent estimates for sources in the reionisation era \citep[e.g.][]{Stark2017,Endsley2021}. For various relative strengths of the burst and the evolved populations, we compute the {\it JWST}/NIRCam photometry for this mock $z=8$ galaxy using the NIRCam wide and medium filter transmission curves ensuring a SNR $=10$ to evaluate the uncertainties (i.e., the SNR that NIRCam reaches to observe a point source with M$_{\rm{UV}}\simeq-18$ with $t\simeq10$~ks; \citealt{Robertson2021}). Using the BAGPIPES nonparametric SFH models described in Section~\ref{sec:bagpipes}, we attempt to detect the underlying evolved stellar population.

We find that even when the evolved stellar mass is $10\times$ the burst mass, nonparametric models cannot convincingly detect the presence of an evolved stellar population for the most extreme line emitters. This is the case for a system where we fix the burst ($5$~Myr) stellar mass to $10^8\ M_{\odot}$ and the evolved stellar mass to $10^9\ M_{\odot}$. Although the nonparametric models can adequately recover the mass formed in the burst phase ($M_{\star}=(1.1\pm0.2)\times10^8\ M_{\odot}$ in the $0-3$~Myr age bin), the stellar mass formed at $>100$~Myr is significantly underestimated (median $M_{\star}=1.8\times10^8\ M_{\odot}$) with a large uncertainty ($1\sigma$ range $4\times10^3-1\times10^9\ M_{\odot}$). When the evolved stellar mass is $20\times$ the burst mass, it is more readily revealed (median $M_{\star}=1.5\times10^9\ M_{\odot}$) but the uncertainty remains large ($1\sigma$ range $6\times10^3-3\times10^9\ M_{\odot}$). Here we notice that by choosing a different prior for the stellar mass distribution in the time bins in nonparametric SFH modeling (Section~\ref{sec:bagpipes}) might lead to a different median stellar mass. For example, a Dirichlet prior distribution favours an older mass-weighted stellar age or a longer star formation timescale \citep{Leja2019,Tacchella2022a} than the uniform logarithm mass prior we used. Thus, we do not rule out that the choice of a different prior could potentially result in a derived median mass that was closer to the mass of the evolved population. However, without the knowledge of rest-frame NIR luminosity, it is difficult to robustly constrain the true stellar mass with small uncertainties.

The simulation described above reveals the large uncertainties associated with inferring the assembly history of galaxies in the reionisation era from such intense line emitters. Recent studies of $z\sim7-8$ star-forming galaxies have argued that many are young systems with relatively low stellar masses \citep[e.g.][]{Labbe2013,Stefanon2022}. These conclusions are usually derived by fitting the {\it HST} and {\it Spitzer} SEDs (i.e., rest-frame UV and optical at $z\sim7-8$) with parametric SFH models (e.g., constant SFR). However, we have demonstrated in Section~\ref{sec:old_star} that the stellar masses of EELGs could be underestimated by a factor of $\sim40\times$ when considering CSFH fitting due to the difficulty of locating evolved stellar populations. Although it is perfectly possible that EELGs at $z\sim1-3$ may not share the same SFHs as those at $z\sim7-8$, our nonparametric fitting of mock NIRCam SEDs at $z=8$ suggests that the stellar masses of the most extreme line emitters could still be underestimated by $\sim10\times$ if they are derived from rest-frame UV and optical photometry. Evidence of evolved stars has already been identified in a handful of galaxies at $z\gtrsim9$ which formed prior to $z\simeq12$ \citep[e.g.][]{Hashimoto2018,Roberts-Borsani2020,Laporte2021}. As shown in the simulation at $z=8$ and the results inferred from EELGs at $z\sim1-3$, if the stellar masses of the most extreme [O~{\small III}] emitters (EW$_{\rm{[OIII]}\lambda5007}>800$~\AA), which compose $20$~per~cent of the $z\sim7-8$ population \citep{Endsley2021}, were underestimated by a factor $\sim10-40\times$, the total stellar mass density at $z\sim7-8$ could be underestimated by a factor $\sim2-8\times$. Conservatively, it seems reasonable to assume the mass density is underestimated by at least a factor $2\times$.

Finally, we consider the cosmic evolution of the UV luminosity density (and hence the SFR density) in the reionisation era in the context of the SFHs of EELGs presented in this study. \citet{Oesch2018} have argued for a rapid decline of the UV luminosity density at $z>8$, while \citet{McLeod2016} suggested a smoother decline. Here we revisit the test provided in the discussion sections of \citet{Roberts-Borsani2020} and \citet{Laporte2021}. Considering the population of $z=8$ galaxies, we examine the fraction of their stellar mass that formed at earlier times. We focus on the stellar mass that formed at $z>9$, which represents an age $>100$~Myr for sources viewed at $z=8$. To derive the fraction of stellar mass formed at $z>9$ relative to $z=8$, we adopt the cosmic evolution of the SFR density in \citet{McLeod2016} and \citet{Oesch2018}, which is converted from the UV luminosity density and assuming zero dust attenuation at $z>8$, and integrate the SFR with time to compute the stellar mass formed at a given redshift. 

\begin{figure}
\begin{center}
\includegraphics[width=\linewidth]{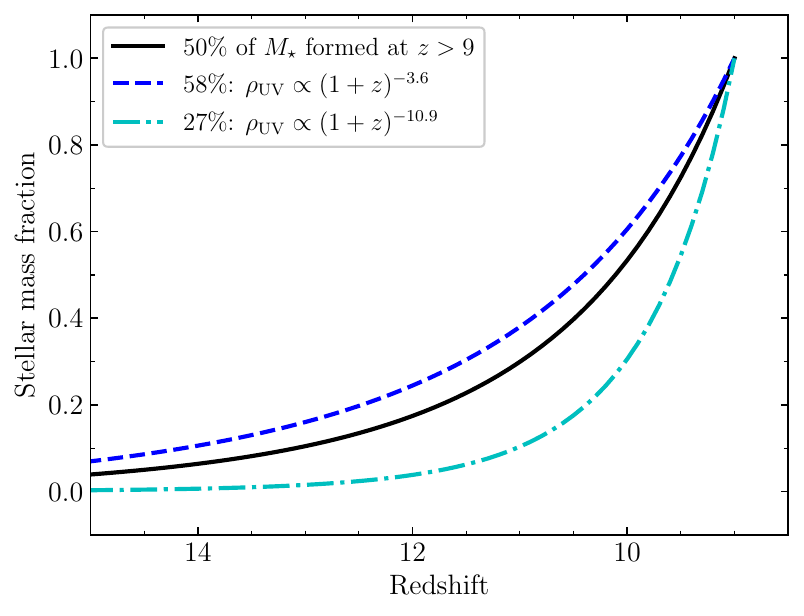}
\caption{Redshift evolution of the fraction of stellar mass in $z=8$ galaxies that formed at $z>9$ (i.e., with an age $>100$~Myr). A smooth decline of UV luminosity function at $z>8$ ($\rho_{\rm{UV}}\propto[1+z]^{-3.6}$, blue dashed line; \citealt{McLeod2016}) would imply $58$~per~cent of the stellar mass at $z=8$ was already in place at $z>9$, whereas this fraction would only be $27$~per~cent in the case of a rapid decline ($\rho_{\rm{UV}}\propto[1+z]^{-10.9}$, cyan dash-dotted line; \citealt{Oesch2018}). As demonstrated in our simulation of fitting NIRCam SEDs at $z=8$ with nonparametric SFH models, the stellar masses of $20$~per~cent of the galaxies at $z=8$ (i.e., the most extreme line emitters; \citealt{Endsley2021}) could be underestimated by $10\times$ if we only probe the rest-frame UV and optical photometry. Thus, the total stellar mass density at $z=8$ could be underestimated by $2\times$, that $50$~per~cent of the stellar mass was formed at $z>9$ (black solid line, averaged at $z>9$), which favours a smooth decline. All the three curves are normalized at $z=9$.}
\label{fig:mass_frac}
\end{center}
\end{figure}

In Fig.~\ref{fig:mass_frac}, we show the redshift evolution of the fraction of stellar mass formed relative to $z=8$. Adopting the power-law function proposed by \citet{Oesch2014}, a rapid decline of the UV luminosity density ($\rho_{\rm{UV}}\propto(1+z)^{-10.9}$) indicates that $27$~per~cent of the stellar mass in $z=8$ galaxies was formed at $z>9$ (cyan dash-dotted line in Fig.~\ref{fig:mass_frac}), while this fraction becomes $58$~per~cent in the case of smooth decline ($\rho_{\rm{UV}}\propto(1+z)^{-3.6}$; blue dashed line in Fig.~\ref{fig:mass_frac}). As demonstrated in our simulation of fitting NIRCam SEDs at $z=8$ with nonparametric SFH models, if the stellar masses of the most extreme line emitters (which compose $20$~per~cent of the total population at $z=8$; \citealt{Endsley2021}) were underestimated by $10\times$, the total stellar mass density at $z=8$ could be underestimated by $2\times$. In this case, about $50$~per~cent of the stellar mass at $z=8$ would be formed at $z>9$, which is consistent with a smooth decline of the UV luminosity density at $z>8$ (black solid line in Fig~\ref{fig:mass_frac}). Eventually, {\it JWST} observations with MIRI, which is capable of probing rest-frame NIR photometry at $z>7$, or deep NIRSpec observations targeting age indicators such as Balmer absorption lines, could help to determine the age and the assembly history of those systems in the reionisation era dominated by the light of very young stellar populations. 

%%%% Figure: Fraction of stellar mass formed at z>9 %%%%

%%%%%%%%%%%% ACKNOWLEDGEMENT %%%%%%%%%%%%

\section*{Acknowledgement}

MT and RSE acknowledge funding from the European Research Council under the European Union Horizon 2020 research and innovation programme (grant agreement No. 669253). DPS acknowledges support from the National Science Foundation through the grant AST-2109066. The authors thank Lily Whitler for helpful conversations about non-parametric star formation history modeling. We also thank St\'{e}phane Charlot and Jacopo Chevallard for providing access to the BEAGLE SED fitting code.

This work is based on observations taken by the 3D-HST Treasury Program (GO 12177 and 12328) with the NASA/ESA HST, which is operated by the Association of Universities for Research in Astronomy, Inc., under NASA contract NAS5-26555. Some of the data presented herein were obtained at the W. M. Keck Observatory, which is operated as a scientific partnership among the California Institute of Technology, the University of California and the National Aeronautics and Space Administration. The Observatory was made possible by the generous financial support of the W. M. Keck Foundation. The authors wish to recognise and acknowledge the very significant cultural role and reverence that the summit of Maunakea has always had within the indigenous Hawaiian community. We are most fortunate to have the opportunity to conduct observations from this mountain. Part of the observations reported here were obtained at the MMT Observatory, a joint facility of the University of Arizona and the Smithsonian Institution. We acknowledge the MMT queue observers for assisting with MMT/MMIRS observations.

This research made use of {\footnotesize ASTROPY}, a community-developed core {\footnotesize PYTHON} package for Astronomy \citep{AstropyCollaboration2013}, {\footnotesize NUMPY}, {\footnotesize SCIPY} \citep{Jones2001}, and {\footnotesize MATPLOTLIB} \citep{Hunter2007}.

%%%%%%%%%%%% DATA AVAILABILITY %%%%%%%%%%%%

\section*{Data Availability}

The data underlying this article will be shared on reasonable request to the corresponding author.



\bibliographystyle{mnras}
\bibliography{stellar_population}

%%%%%%%%%%%%%%%%%%%%%%%%%%%%%%%%%%%%%%%%%%%%%%%%%%

%%%%%%%%%%%%%%%%% APPENDICES %%%%%%%%%%%%%%%%%%%%%

\appendix

%%%%%%%%%%%%%%%%%%%%%%%%%%%%%%%%%%%%%%%%%%%%%%%%%%

% Don't change these lines
\bsp	% typesetting comment
\label{lastpage}
\end{document}